\documentclass[letterpaper]{ar-1col}
\usepackage{aas_macros}
\usepackage{rotating}%
\usepackage{natbib}%
\usepackage{amsfonts,amsmath,amssymb}
\usepackage{floatrow}%
\usepackage{subfigure}%
\usepackage[export]{adjustbox}
\usepackage[margin=1.5in]{geometry}
\jname{Xxxx. Xxx. Xxx. Xxx.}
\jvol{00}
\jyear{YYYY}
\doi{10.1146/((please add article doi))}
\usepackage{graphicx}
\usepackage{epstopdf}
\DeclareMathSymbol{\varOmega}{\mathord}{letters}{"0A}
\DeclareMathSymbol{\varSigma}{\mathord}{letters}{"06}

\DeclareMathSymbol{\varPsi}{\mathord}{letters}{"09}

\newcommand{\de}{\mbox{d}}

\newcommand{\dln}{\left|\frac{\de\ln\Omega}{\de\ln R}\right|}

\newcommand{\Eq}[1]{Equation\,(\ref{#1})}

\newcommand{\Fig}[1]{Figure~\ref{#1}}

\newcommand{\eqfrac}[2]{\left(\frac{#1}{#2}\right)}

 \newcommand{\Mdot}{\dot{M}}
\newcommand{\Msun}{M_{\odot}} 
\newcommand{\Lsun}{L_{\odot}}

\newcommand{\Mjup}{M_{\rm Jup}}
\newcommand{\lsim}{\mathrel{\rlap{\lower4pt\hbox{\hskip1pt$\sim$}}
    \raise1pt\hbox{$<$}}}                
\newcommand{\gsim}{\mathrel{\rlap{\lower4pt\hbox{\hskip1pt$\sim$}}
    \raise1pt\hbox{$>$}}}                

\newenvironment{dedication}
  {\clearpage           
   \thispagestyle{empty}
   \vspace*{\stretch{1}}
   \itshape             
   \raggedleft          
  }
  {\par 
   \vspace{\stretch{3}} 
   \clearpage           
  }
\begin{document}

\markboth{Kratter \& Lodato}{GI DISKS}

\title{Gravitational Instabilities in Circumstellar Disks}

\author{Kaitlin M. Kratter$^1$ \& Giuseppe Lodato$^2$ 
\affil{$^1$Department of Astronomy and Steward Observatory, Univ. of Arizona, 933 N Cherry Ave, Tucson, AZ, 85721}
\affil{$^2$Dipartimento di Fisica, Universit\`a degli Studi di Milan, Via Celoria 16, 20133 Milan, Italy}
}


\begin{abstract}
Star and planet formation are the complex outcomes of gravitational collapse and angular momentum transport mediated by protostellar and protoplanetary disks. In this review we focus on the role of gravitational instability in this process. We begin with a brief overview of the observational evidence for massive disks that might be subject to gravitational instability, and then highlight the diverse ways in which the instability manifests itself in protostellar and protoplanetary disks: the generation of spiral arms, small scale turbulence-like density fluctuations, and fragmentation of the disk itself. We present the analytic theory that describes the linear growth phase of the instability,  supplemented with a survey of numerical simulations that aim to capture the non-linear evolution. We emphasize the role of thermodynamics and large scale infall in controlling the outcome of the instability. Despite apparent controversies in the literature, we show a remarkable level of agreement between analytic predictions and numerical results. In the next part of our review, we focus on the astrophysical consequences of the instability. We show that the disks most likely to be gravitationally unstable are young and relatively massive compared to their host star, $M_{\rm d}/M_* \geq 0.1$. They will develop quasi-stable spiral arms that process infall from the background cloud.  While instability is less likely at later times, once infall becomes less important, the manifestations of the instability are more varied. In this regime, the disk thermodynamics, often regulated by stellar irradiation, dictates the development and evolution of the instability. In some cases the instability may lead to fragmentation into bound companions. These companions are more likely to be brown dwarfs or stars than planetary mass objects. Finally, we highlight open questions related to (1) the development of a turbulent cascade in thin disks, and (2) the role of mode-mode coupling in setting the maximum angular momentum transport rate in thick disks. 

\end{abstract}


\maketitle
\begin{dedication}
{We dedicate this article to the memory of Francesco Palla, who has been a source of inspiration for star formation and disk studies throughout the years. Through his support and encouragement of young scientists, his contributions live on.}
\end{dedication}

\section{INTRODUCTION}

Protostellar and protoplanetary disks stand at the center of our study of origins. They regulate the formation of stars and the planetary systems that they host.  Over the last three decades, the observational study of protostellar disks has evolved from quantifying hints of infrared excess \citep{Kenyon:1987} to resolving ring-like structures at 1AU scales \citep{hltau}. Long wavelength studies with facilities such as the SMA, Plateau de Bure, CARMA, and now ALMA, combined with sensitive infrared measurements from IRAS and Spitzer enable detailed studies of the sizes, temperatures, and masses of a variety of disks in young clusters. While many uncertainties remain in translating high resolution interferometric measurements into reliable disk properties, we can begin to see the various stages of star and planet formation occurring in these disks.

Theoretical studies have made similar progress, from simple, 1-D diffusive models to three-dimensional (magneto)hydrodynamics simulations that incorporate radiative transfer. These advances have been enabled by the burgeoning computational resources at our disposal. While the most complex of these models provide many insights, which we review in detail, simple models prove valuable to this day.

In this review we discuss the role that self-gravity plays in regulating the growth and evolution of protostellar and protoplanetary disks. Self-gravity in general refers to the mutual gravitational influence of relatively diffuse gas on itself, in contrast to the gravity exerted on the gas by an external body. In this context, self-gravity refers to the effect of the gravitational field of the disk itself, rather than the protostar, on the evolution of the system.  Although the dividing line between protostellar and protoplanetary disks is fuzzy, we use the former to refer to younger disks surrounding protostars that are not yet near their final mass, while the latter refers to the remnant disk in which the planet formation process completes (of course it may begin much earlier).  The physics we discuss is relevant for all stellar masses, but much of our scalings are for sun like stars.

We focus our attention on hydrodynamics, and  thus we do not discuss the important role that magnetic fields and MHD effects have on the evolution of protostellar disks. It is well known that the hotter parts of protostellar disks are subject to the magneto-rotational instability \citep{balbusreview}, that can provide the bulk of angular momentum transport. On the other hand, the colder, outer disks, which are the regions most likely to become gravitationally unstable, are probably less affected by MHD turbulence, given their low ionization state. Very few numerical studies have considered the combined effects of MHD and gravitational instabilities in disks \citep{fromang04a,fromang04b}, whose interplay might be important in protostellar disk evolution.

In Section \ref{sec-obsev} we begin with a brief observational overview with an eye towards the applicability of the physics of self-gravitating fluids to observed systems. In Section \ref{sec-sgphys} we provide an in-depth review of the nature of self-gravity and gravitational instabilities, including the non-linear outcome of the instability: fragmentation of a disk. The physics discussed here is applicable to accretion disks in many different astrophysical contexts. In Section \ref{sec-sgprotostellar} we hone in on disks around young stars, and consider how they fit into the various regimes outlined in Section \ref{sec-sgphys}. In Section \ref{sec-sgplanet} we discuss the relevance of gravitational instabilities to the formation of planets both directly (gas-collapse) and indirectly (concentration of dust particles). We conclude with a brief summary and prospects for future.

\section{Observational Evidence for Self-Gravitating Disks} \label{sec-obsev}
Before reviewing the theoretical underpinnings of the role of self-gravity in protostellar and protoplanetary disk evolution, we address the observational evidence that this physics is relevant. As we detail in later sections, understanding the behavior of self-gravitating disks is complex and depends on a range of disk properties including size, surface density, temperature, and thermal physics. The standard reference for quantifying the degree to which a disk is self-gravitating is the parameter $Q$, defined as \citep{Toomre:1964}:
 \begin{equation}\label{eq-toomreQ}
Q = \frac{c_s\kappa}{\pi G \Sigma}
\end{equation}
where $c_s$ is the sounds speed, $\kappa$ is the epicyclic frequency (which is equal to $\Omega$ in a Keplerian disk), and $\Sigma$ is the disk surface density. As $Q\rightarrow 1$ self-gravity becomes increasingly important. We discuss the origin of $Q$ in more detail in section \ref{sec-sgphys}. To translate this constraint into one motivated by observations, we note that $Q$ can be written as:
\begin{equation}\label{eq-mdhr}
Q = \frac{c_s\Omega}{\pi G \Sigma} = f \frac{M_*}{M_{\rm d}}\frac{H}{r}
\end{equation}
where $M_*$ is the stellar mass, $M_{\rm d}$ is the disk mass, $H$ is the disk scaleheight, and $r$ the radius. We subsume numerical factors of order unity, which depend on the surface density profile, into a pre-factor $f$.  While we cannot escape the dependence on the disk aspect ratio, $H/r$, which depends on the temperature, we can use this relation to obtain an order of magnitude estimate for the minimum mass required for a disk to be self-gravitating. Colder, thinner disks are more prone to self gravity. Star forming regions typically do not have temperatures below 10K, so we take that as a lower limit. What about disk radii? Many lines of argument, from measures of cloud angular momentum \citep{Goodman:1993}, to stellar interactions \citep{Adams:2006}, to direct observations in both absorption \citep{Eisner:2008} and millimeter wavelengths \citep{Andrews:2011} suggest that $10^2-10^3$AU are reasonable limits for disk radii. At constant temperature, the aspect ratio increases with disk radius as $r^{1/2}$, so we use 100AU to set $H$ for our scaling. The requirement that $Q<1$ (implying gravitational instability, see below) then implies:
\begin{equation} \label{eq-critdisk}
\frac{M_d}{M_*} > 0.06\left(\frac{f}{1}\right)\left(\frac{T}{10K}\right)^{1/2}\left(\frac{r}{100AU}\right)^{1/2}\left(\frac{\Msun}{M_*}\right)^{1/2}
\end{equation}
Thus for self-gravity to be important, i.e. for $Q$ to be near unity given optimistic constraints on temperature, we expect disk-star mass ratios greater than $\approx10^{-2}$. 

With this benchmark in mind, we now consider the evidence for disks with mass ratios in this regime or higher.

\subsection{The Phases of Disks}
Early studies of protostellar disks in the infrared engendered a classification system based on the slopes of disk+star spectral energy distributions (SEDs). Three classes, 0, I, and II, correspond to positive, flat, and negative power law indices longward of 2 microns.  
In addition there is a final phase, Class III, wherein the SED is dominated by the stellar photosphere, but weak Infrared excess indicates remaining disk material.
There is some evidence that this sequence corresponds with evolutionary states. Class 0 protostellar sources correspond to young, deeply embedded objects, whose disk-like component is hard to measure. Class I sources are thought to have similar mass in the envelope and disk, and Class II sources have mostly finished accreting from their natal envelope, revealing just the protoplanetary disk, as most of the stellar accretion has been completed.  Class III disks are not thought to be self-gravitating.

\subsection{Measuring Disk Masses}
Measuring disk masses at any evolutionary state remains extremely challenging. Even as observatories like ALMA grow ever more sensitive, several issues limit precise mass measurements.  Many limitations are related to the fact that most disk masses are measured by proxy through detection of millimeter wavelength fluxes associated with dust grains of up to a millimeter in size. Based on studies of the ISM, the ratio of gas to dust is thought to be of order $10^2$, but this is not well confirmed in disks. Secondly, because larger grains and rocks do not emit at millimeter wavelengths, these data do not reveal the potentially sizeable fraction of dust which has already grown, something for which there is already strong evidence (e.g.,  \citealt{Perez:2012}).  Moreover, the conversion of a flux to a $<$mm-size dust mass requires adopting an opacity, which is itself dependent on unknowns like the grain-size distribution, and grain properties such as chemical composition and porosity (see, e. g. \citealt{natta04,ricci10}). Finally, mass measurements from dust are only possible if the dust is optically thin. If the dust is optically thick, which is possible in the inner regions of disks (soon to be resolved by ALMA) only a lower limit on the mass is possible. 

In addition to the uncertainties generated by the dust properties, \cite{Dunham:2014} have highlighted the estimation errors introduced by contamination of disk flux measurements due to the presence of a core at early phases, leading to overestimated masses at early times. Similarly, as disks grow to larger radii, high resolution interferometric studies can also resolve out large scale structures where much of the mass resides. This error leads to a more severe underestimate of disk masses.

With these caveats in mind, we review the best estimates of disk masses.

\subsubsection{Observational Results}
The most comprehensive survey for disk masses at this time is that of \cite{Andrews:2013}, who survey all known members of the nearby star forming region Taurus.  These disks are mostly Class II sources, however some class I sources are included. 

\citet{Andrews:2013} compiles photometry at $1.3mm$ using the SMA to derive both disk masses, and disk-star-mass ratios.
The stellar masses are estimated using the isochrones from \citet{Dantona:1997,Baraffe:2002} and \citet{Siess:2000}. Disk masses are estimated from the submillimeter flux assuming a constant distance of $140$pc, dust-gas ratio $\zeta = 0.01$, dust opacity: $\kappa_\nu = 2.3 \rm{cm}^2/\rm{g}$ and a disk temperature constant with radius, which scales with stellar luminosity as $\langle T_d\rangle = 25(L_*/\Lsun)^{1/4}$K. The masses are derived from the following equation for optically thin, isothermal dust:
\begin{equation}
\log M_{\rm d} = \log F_\nu + 2 \log d-\log(\zeta \cdot \kappa_\nu) - \log B_\nu(\langle T_d\rangle),
\end{equation}
where $F_\nu$ is the observed flux, $B_\nu$ is the Planck function and $d$ is the distance to the source.
 
It is apparent from \cite{Andrews:2013} figure 11 that most disk masses are far too low in mass by 2-3 Myr (predominantly class II stage) for GI to be relevant. Note additionally that the assumed outer disk temperatures for this sample are higher than our optimistic (for triggering GI) estimate of 10K. Depending on order unity factors, at most 20\%, and probably more like 10\% of Class II disks are plausibly massive enough for self-gravity to be important. A self-consistent treatment of the disk temperatures in our instability threshold (\Eq{eq-critdisk}) and mass estimates would yield critical mass ratios closer to $0.1$, higher than nearly every disk in the sample.

As this sample is dominated by Class II disks, we may conclude that both angular momentum transport and planet formation reliant on strong self-gravity are rare at best for systems which have evolved to the Class II stage. 

What about evidence for younger, massive Class I and Class 0 disks? Unfortunately at the time of this writing the evidence remains sparse. Class 0 and I sources are deeply embedded and thus intrinsically harder to both identify and measure. \cite{Eisner:2012} and \cite{Sheehan:2014} have isolated and modelled in detail the Class I sources in Taurus. They find that the masses are typically higher than in the Class II phase by a factor of two: the median mass of Class I disks is $0.008\Msun$ compared to $0.005\Msun$ for older sources. Among Class I sources,  roughly 10\% have masses higher than $0.01\Msun$ \citep{Eisner:2012}, at the margins of susceptibility to strong self-gravitating effects. Some individual sources have been measured to have relatively massive and compact disks in the Class 0/I phase, for example IRAS 16293-2422B \citep{rodriguez05}, or WL12 \citep{miotello15}.

A recent survey by \cite{Mann:2015} also hints that younger sources may indeed have more massive disks. These authors find that in NGC2024 (thought to be $<1$Myr old as compared to 2-3 Myr for Taurus) as many as 10\% of disks have masses greater than $0.1\Msun$, with 20\% above $0.01\Msun$. Despite the $2\sigma$ significance of higher mass measurements, selection biases alone (related to stellar mass and multiplicity) could account for the increase if the cluster properties of Taurus and NGC2024 are similar. 

Finally, we turn to observations of disks in the Class 0 stage. The most conclusive evidence for a relatively massive disk comes from \cite{Tobin:2013}. Using the SMA and CARMA, they detect an edge on $\approx 150-200AU$ disk surrounding the Class 0 protostar L1527. Using a similar optically thin dust model they derive a mass comparable to Class I sources of $0.075\Msun$, however the protostellar mass is thought to be $<0.5\Msun$, and thus the disk-star mass ratio is easily high enough for self-gravity to be important. There are several other tentative detections of Class 0 disks (see e.g. \citealt{Looney:2000,Enoch09}). A larger survey in Perseus by \cite{Tobin:2015} resolved several more likely candidates and found that more than 50\% of 9 sources were more consistent with envelope + disk models rather than pure envelope models. Additionally several sources show evidence for rotation (though not necessarily Keplerian motion). The masses measured in this survey cannot reliably separate out all envelope contamination, thus the increase in masses (up to $0.1\Msun$) is not necessarily indicative of strong self-gravity. Nevertheless it is quite plausible that disk masses are higher in the Class 0 phase.

\cite{Maury:2010} have made strong claims against the presence of disks in the Class 0 phase, but the well-resolved L1527 was included in their sample, suggesting that perhaps the data was insufficient to rule out some extended disks. \cite{Tobin:2013} had slightly better resolution, and also a different beam alignment relative to the disk direction, which might explain the \cite{Maury:2010} non-detection.

While dust mass measurements are most common, recent efforts have been made to directly detect molecular gas tracers to constrain the total mass and dust-to-gas ratio. \cite{Bergin:2013} use Herschel data to detect HD in TW Hya, finding a relatively large disk mass for a nominally 10 Myr system. \cite{Williams:2014} used optically thick and thin CO isotopologues and conclude that most Class II disks have dust-to-gas ratios up to a factor of 10 below ISM values, implying that either planet formation is well underway, or total disk masses are even smaller.

The above discussion focused on solar type and Herbig AeBe stars, for which there is more data, but there is some evidence that high mass stars host massive, self-gravitating disks (\citealt{cesaroni07} and references therein). Precise measurements remain elusive because high mass stars are typically further away, and thus harder to resolve. Additionally, the only evidence for disks coincides with the equivalent of the Class 0-I phases, when the stars are deeply embedded in a gas envelope. For example, rotationally supported disks are difficult to distinguish from rotationally flattened envelopes \citep{Johnston:2011}. One of the more convincing examples \citep{Shep2001,Shepherd:2004} reveals a rotating massive disk around a B-star that is likely gravitationally unstable based on the estimated parameters. \cite{krumholz07} has shown that a fully functional ALMA may be able to validate these observations using molecular lines.

\subsection{Direct Evidence for Disk Self-Gravity}
While the population at large may not show strong evidence for disk self-gravity, a handful of individual sources do show indications of the spiral arm structure typically associated with spiral density waves. These wave may be triggered either by a young planet, or by disk self-gravity. \cite{Muto:2012} and \cite{Wagner:2015} find evidence for spiral structure in scattered light images of transition disks. Self-gravity might be at play, but the low sub-mm masses make this questionable. \cite{Benisty:2015} also find evidence for spiral structures in scattered light images, but claim a planetary origin. ALMA data might settle these two theories by providing disk mass constraints, but at present neither theory can be ruled out \citep{dong15}.

\subsection{Indirect Evidence for Disk Self-Gravity}
While observational examples of disks with strong self-gravity remain scarce, there are other indirect sources of evidence from comprehensive studies of star formation. There are a variety of reasons to expect that disks may be more massive at earlier evolutionary states. First, since the discovery of the first statistically significant sample of protostars, observers and theorists alike have pointed out the ``luminosity problem" \citep{Hartmann:1996}. This problem arose due to the mismatch between the observed protostellar luminosities, of the order of a few $L_{\odot}$, and the expected luminosities derived from basic energetic arguments that predict that the accretion luminosity:
\begin{equation}
L_{\rm{acc}} \approx \frac{GM_*\Mdot}{R_{\rm in}} \approx 100 \Lsun \frac{M}{\Msun}\frac{\Mdot}{10^{-6}\Msun/yr}\frac{R_{\rm in}}{4R_\odot}
\end{equation}
where $R_{\rm in}$ is the inner disk radius and $\dot{M}$ is the infall rate, estimated based on a star formation timescale of roughly 1Myr. This over prediction of the typical luminosity has been used as evidence for large, short-lived accretion events, either bursts, or more continuous but rapid events early in a protostars history. A variety of theories have been proposed for producing these bursts. 

Before considering specific theories, we note the following simple scaling argument. Consider a star forming core or turbulent filament. Observational evidence suggest that these bodies eventually undergo a modified form of free fall collapse when self-gravity overcomes magnetic or turbulent pressures. Thus the rate at which material will fall onto any disk like structure formed in the collapse is dictated by free fall collapse. The oversimplified isothermal inside out collapse mode of \cite{Shu1977} provides an order of magnitude estimate for this infall rate:
\begin{equation}
\Mdot_{\rm in} \approx \frac{c_s^3}{G} \approx 1.6 \times 10^{-6} \left(\frac{T}{10K}\right)^{3/2} \Msun/yr.
\end{equation}
Somewhat more realistic Bonnor Ebert sphere infall rates are the same order of magnitude, and are consistent with the timescale on which star formation occurs. In contrast, the observed accretion rates from disks on to stars in the Class I - Class II phase range from $10^{-9}-10^{-7}\Msun/\rm{yr}$, with the majority of measurements at the lower end \citep{Gullbring:1998}.  There are two possible means to resolve the mismatch between infall and stellar accretion rates. Either disks at earlier times are more massive, and thus can process more material at higher rate. This option is somewhat suggestive of self-gravitational effects, but not required. Alternatively if initially disk-star accretion rates are not higher, the mass of the disk will rise, bringing it ever closer to the regime where self-gravity is important. It seems unlikely that the infall and disk-star accretion rates, which rely on entirely different physics (free fall and cloud turbulence vs either self-gravity or magnetic fields), could conspire to match up at all times in order to maintain a constant, tiny fraction of the total mass in the disk. Most of the stellar mass is thought to pass through the disk at some point, making such a balance improbable.

\cite{Armitage:2001} first suggested that stellar accretion and FU Orionis type outbursting systems might be generated by an interplay between gravitational instability driven accretion and magnetically dominated accretion driven by MRI turbulence.  \cite{VorBas06} suggest that these same outbursts could be triggered by clumps formed due to gravitational instability in the outer disk that are tidally disrupted as they migrate inwards, and subsequently accrete onto the star. See the chapter by Hartmann in this Annual Reviews (and recently, the Protostars and Planet VI chapter, \citealt{audard2014}) for more details on variable accretion onto pre-main-sequence stars.

One final piece of evidence for higher disk masses relies on observations of exoplanets. \cite{Najita:2014} have shown that typical protoplanetary disks do not have enough observed dust mass to account for the mass of planets implied by  occurence rates derived from {\em Kepler}. They infer that planet formation, and thus grain growth likely commenced at even earlier times. This implies that the observed dust masses are missing a substantial amount of solids. Accounting for this missing mass might well push a larger fraction of disks into the self-gravitating range.

In conclusion we see that while inferred disk masses are often lower than the nominal requirement for strong self-gravity, it is by no means ruled out. Given the many ways to underestimate disk masses, the observational evidence demands the study of disk self-gravity .

\section{Physics of self-gravity and gravitational instability}
\label{sec-sgphys}
\begin{marginnote}
\entry{GI}{ Gravitational Instability}
\entry{MRI}{Magneto-Rotational Instability}
\entry{SPH}{Smoothed Particle Hydrodynamics}
\end{marginnote}
In the previous section we have shown that disks may be subject to the effects of self-gravity when the ratio $M_{\rm d}/M_* > 10^{-2}$. We first address how self-gravity affects the basic hydrostatic structure of the disk, and subsequently explore how gravitational instability, hereafter GI, may affect disk structure and evolution. In this section we focus on the physical nature of self-gravity and GI, which applies to disks in a wide range of astrophysical contexts. We begin by discussing isolated disks with a fixed surface density profile for simplicity, and subsequently consider the changes that arise in the presence of infall and external heating. In Section \ref{sec-sgprotostellar}, we discuss the extent to which such physical processes apply to actual protostellar disks.

\subsection{Effects of self-gravity on the structure of accretion disks}
Consider a thin axisymmetric disk surrounding a protostar of mass $M_{\star}$. Let $\rho(r,z)$, $\Sigma(r)$ and $c_{\rm s}(r)$ be the volume density, the surface density and sound speed of the gas as a function of cylindrical radius $r$, and height above the disk midplane $z$, respectively. Standard, non self-gravitating disk models are based on the assumptions of centrifugal balance in the radial direction and hydrostatic balance in the vertical direction. Both equations must be modified to account for self-gravity. 

A standard non self-gravitating disk in hydrostatic balance is characterised by a Gaussian profile for $\rho$, where the disk scale height is given by 
\begin{equation}
H_{\rm nsg}=\frac{c_{\rm s}}{\Omega_{\rm k}},
\end{equation} 
where $\Omega_{\rm k}=\sqrt{GM_*/r^3}$ is the Keplerian angular velocity and we have assumed the disk to be vertically isothermal. Self-gravity modifies vertical hydrostatic balance. In the limit where self-gravity dominates, the vertical density profile is given by \citep{spitzer42}
\begin{equation}
\rho(z)=\frac{\rho_0}{\cosh^2(z/H_{\rm sg})},
\end{equation}
where 
\begin{equation}
H_{\rm sg}=\frac{c_{\rm s}^2}{\pi G\Sigma}.
\end{equation}
In order to neglect self-gravity in the vertical direction, we require that $H_{\rm nsg}/H_{\rm sg}\ll 1$. This immediately translates into the condition
\begin{equation}
\frac{c_{\rm s}\Omega_{\rm k}}{\pi G \Sigma}\gg 1.
\label{eq:toomre1}
\end{equation}
From this requirement alone we thus recover the usual stability parameter for GI (\citealt{Toomre:1964}). This means that whenever self-gravity is important in the vertical direction, the disk is near the instability threshold for GI. This constraint is also equivalent to the constraint on disk mass presented in \Eq{eq-mdhr} in Section \ref{sec-obsev}.
In the general case where both contributions (of the central star and of the disk) are important, there is no analytical solution to the hydrostatic balance equation, but a simple and accurate interpolation formula between the two results above has been found by \citet{BL99} (see also \citealt{lodatoNC,Munoz:2015}):
\begin{equation}
H=\frac{c_{\rm s}^2}{\pi G\Sigma}\left(\frac{\pi}{4Q^2}\right)\left[\sqrt{1+\frac{8Q^2}{\pi}}-1\right]=\frac{c_{\rm s}^2}{\pi G\Sigma}f(Q).
\label{eq:thicknessBL}
\end{equation} 

    We now consider centrifugal balance. To lowest order, centrifugal balance prescribes the angular velocity of the disk $\Omega$ to be Keplerian, $\Omega(r)=\sqrt{GM_{\star}/r^3}$. Even in the absence of self-gravity, a small correction is required to account for pressure gradients, which cause the gas to orbit at a slightly sub-Keplerian speed. This sub-Keplerian motion has a dramatic effect on the dynamics of small solids in the disk, whose orbits are Keplerian due to the absence of pressure forces. Taking these corrections into account, we have
\begin{equation}
\Omega(r)=\Omega_{\rm k}(r)\sqrt{1 + \left(\frac{H}{r}\right)^2\frac{\mbox{d}\ln P}{\mbox{d}\ln r}},
\end{equation}
which gives sub-Keplerian rotation because in general $P$ is a decreasing function of radius. Deviations from Keplerian motion due to the pressure gradient are of order $(H/r)^2$.  Self-gravity also modifies centrifugal balance by a force term which is of the order of $GM_{\rm disk}/r^2$ (a more detailed calculation, obtained by solving Poisson's equation, can be found in \citealt{BL99}). 
Clearly, in order to dominate the rotation curve, the disk mass needs to be implausibly large for protostellar disks. However, it is worth noting that when the disk is marginally gravitationally stable ($Q\approx 1$), the disk mass is of the order of $(H/r)M_{\star}$ (\Eq{eq-mdhr}) and thus the deviation from Keplerian motion due to self-gravity is of the order of $H/r$, stronger than the effect of pressure gradients, though both solids and gas feel the change in $\Omega(r)$ due to disk self-gravity.

\subsection{Dispersion relations}
The most important effect of disk self-gravity on the dynamics of protostellar disks is connected with the development of GI. The linear stability of self-gravitating disks has been investigated for decades in the context of galactic dynamics. In the tightly wound approximation, the WKB dispersion relation for an infinitesimally thin disk is the famous \citet{Lin:1964} relation:
\begin{equation}
(\omega-m\Omega(r))^2=c_{\rm s}^2k^2 - 2\pi G\Sigma|k| + \kappa^2,
\label{eq:linshu}
\end{equation}
where $\omega$ is the wave frequency, $k$ is the radial wavenumber, $m$ is the azimuthal wavenumber (that is, the number of spiral arms produced by the instability), and $\kappa$ is the epicyclic frequency, which, for a Keplerian disk is simply $\Omega$. We now see that the aforementioned $Q$ from \Eq{eq-toomreQ} arises as the  threshold for exponential growth of axisymmetric modes, where stability requires
\begin{equation}
\label{toomre2}
Q=\frac{c_{\rm s}\kappa}{\pi G\Sigma}>1.
\end{equation}

The above dispersion relation and stability criterion have been derived under several important limiting approximations. The first is that the equilibrium structure of the disk is axisymmetric, the second (and most important) is that the disk is infinitesimally thin. It can be shown that finite disk thickness  \citep{vandervoort70,bertinbook2,binney} has the effect of stabilizing the disk by diluting the self-gravity term in Eq. (\ref{eq:linshu}). A simple way to account for this effect is to modifiy eq. (\ref{eq:linshu}) as follows:
\begin{equation}
(\omega-m\Omega(r))^2=c_{\rm s}^2k^2 - 2\pi G\Sigma|k|e^{-|kH|} + \kappa^2,
\label{eq:linshuthick}
\end{equation}
that can be expanded to first order to give
\begin{equation}
(\omega-m\Omega(r))^2=(c_{\rm s}^2 + 2\pi G\Sigma H )k^2 - 2\pi G\Sigma|k| + \kappa^2,
\label{eq:linshuthick2}
\end{equation}
where now the stabilizing effect of disc thickness becomes apparent, as an additional `pressure-like' term in the dispersion relation. By inserting Eq. (\ref{eq:thicknessBL}) in Eq. (\ref{eq:linshuthick2}) we finally get 
\begin{equation}
(\omega-m\Omega(r))^2=(1+2f(Q))c_{\rm s}^2k^2 - 2\pi G\Sigma|k| + \kappa^2.
\label{eq:linshuthick3}
\end{equation}
From \Eq{eq:linshuthick3} we see that the stability of the disk depends only on the single parameter $Q$, and one can show that the marginal stability value decreases to $Q\approx 0.6$, for a Keplerian disk. 

In the context of galactic dynamics, it was recognized many years ago \citep{ostriker73,hohl71,hohl73} that disks that are locally stable according to the $Q$ criterion, might still generate large scale, spiral waves. This is due to the fact that such global modes are not captured by the WKB tightly wound approximation. However, a WKB description of such modes can still be obtained under less restrictive conditions than the tightly wound approximation \citep{laubertin78}. The resulting dispersion relation is more complicated (it is a cubic rather than quadratic expression in $k$) and depends on a new dimensionless parameter $J$:
\begin{equation}
J=m\frac{\pi G\Sigma}{r\kappa^2}\frac{4\Omega}{\kappa}\left|\frac{\mbox{d}\ln\Omega}{\mbox{d}\ln r}\right|^{1/2}\approx \sqrt{6}m\frac{M_{\rm d}}{M_{\star}},
\label{eq:j}
\end{equation}
where the last approximation holds for a Keplerian disk. While the $Q$ parameter balances shear, thermal pressure and disk mass, $J$ is a measure of the disk-star mass ratio. The cubic dispersion relation reduces to the standard Lin and Shu expression for the case of light disks ($J$ or $M_{\rm d}/M_{\star}$ much smaller than unity), while for massive disks, where $J\approx 1$, or $M_{\rm d}/M_{\star}\approx 1/m$) lower $m$, less tightly wound, spiral arms arise.
In both regimes, the most unstable mode scales with $H$. When $H$ is small, the behavior is well captured by a local, even 2D analysis. Otherwise, any manifestation of the instability arises over a significant fraction of the radial extent  of the disk. 
\subsection{The onset of linear instability}
\begin{figure} 
\begin{minipage}{.6\textwidth}
 \centering
 \hspace{-0.5in}
\includegraphics[width=2.4in]{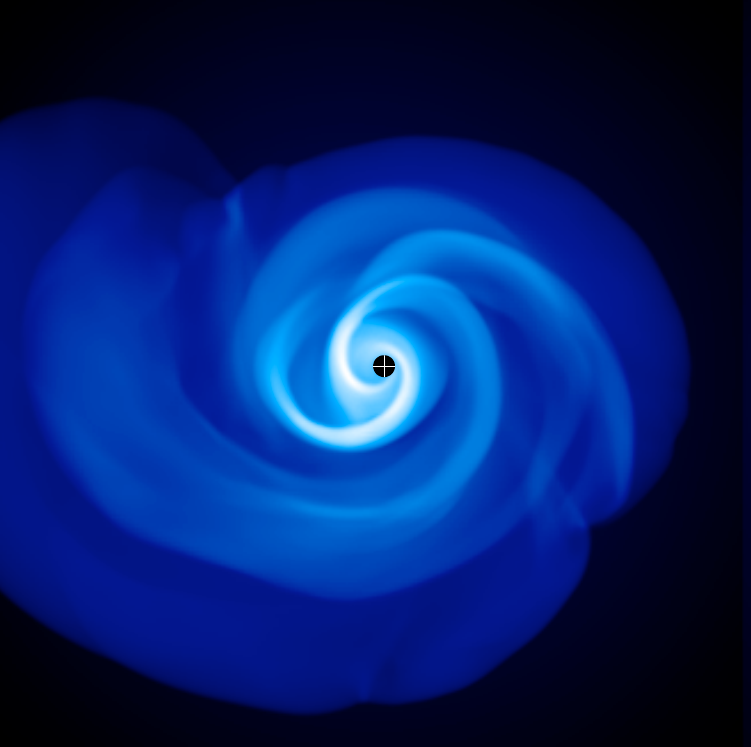}
\end{minipage}%
\begin{minipage}{.6\textwidth}
  \centering
   \hspace{-0.5in}
\includegraphics[width=2.4in]{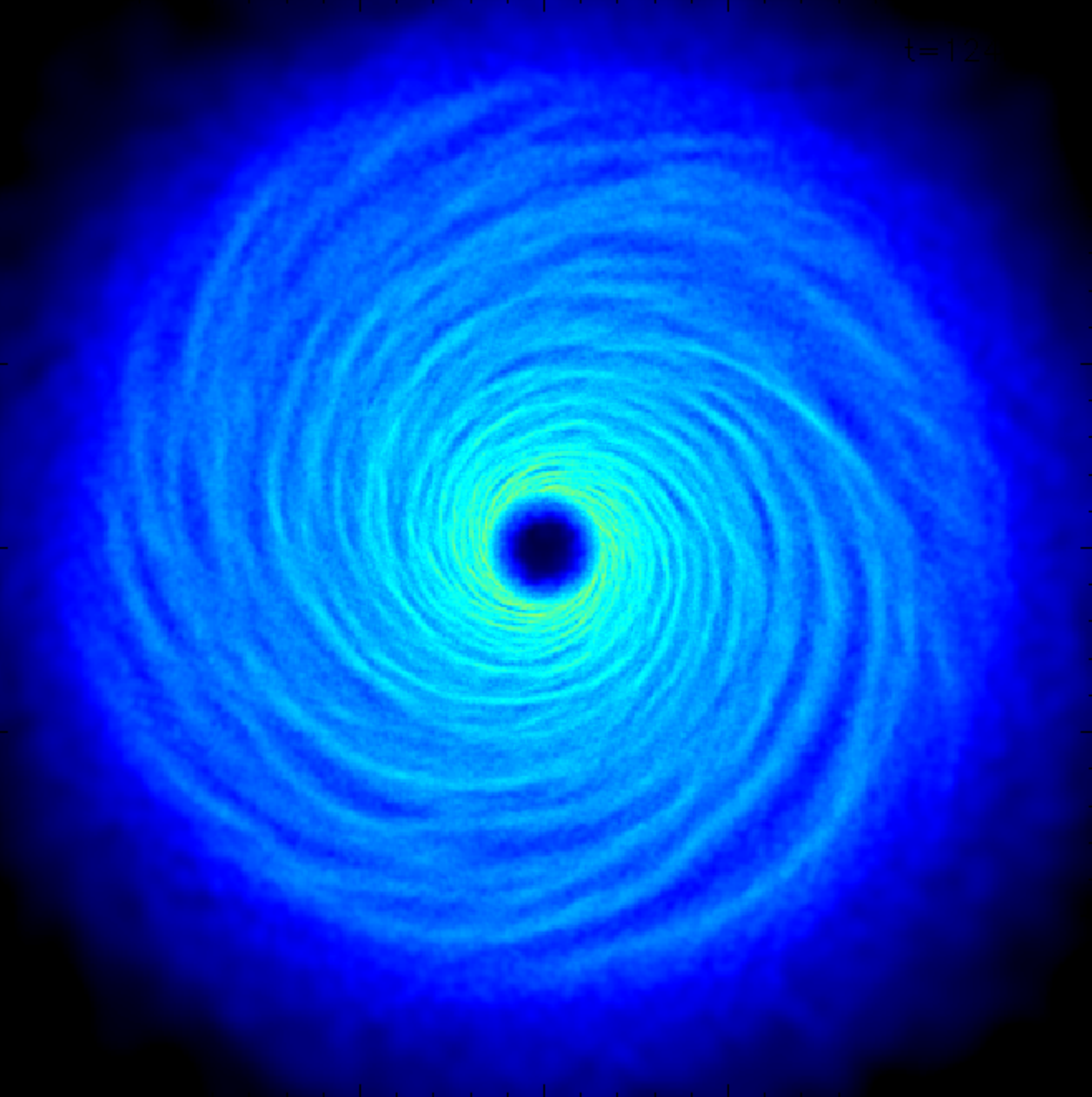}
\end{minipage}%
\caption{At left we show a three-dimensional isothermal simulation from the parameter studies of \cite{KMKK10}, where $M_d/M_* \approx 0.5$. The strong left-right asymmetry is evidence of a dominant $m=1$ mode. At right we show a three-dimensional simulation from \cite{CLC09} with slow cooling, where $M_d/M_* \approx 0.1$. Note the dominance of high $m$ spiral structure.}
\label{fig-sims}
\end{figure}

When the disk becomes linearly unstable, the onset of GI is similar in both regimes of $J$. 
A spectral analysis of GI in numerical simulations confirms that the dominant modes excited have a wavenumber $k\approx 1/H$, as predicted from the Lin \& Shu dispersion relation in \Eq{eq:linshu} \citep{Boley:2007,CLC09,Michael:2012}. Similarly,  the azimuthal wavenumber $m$ scales inversely with $H$ so that thick / massive disks are characterized by fewer spiral arms and by a more open spiral structure.
We provide a more detailed description of the onset of linear instability in each regime below. \Fig{fig-sims} illustrates the morphological differences between the instability with small and large $H/r$.

\paragraph{Tightly Wound, small $H/r$}
The development of small scale instabilities can be captured by both local shearing box / shearing sheet simulations and by thin disk global simulations. This regime is commonly referred to as gravito-turbulence, although some caution should be used when talking about `turbulence' in this context. 
\citet{gammie01} used a Fourier analysis of GI in the shearing sheet approximation to show that the power spectrum of perturbations generated by the instability is contained in modes with wavelengths of a few $H$. \citet{gammie01} finds essentially no power at scales much smaller than $H$, which are resolved in the calculations. A similar result has recently been obtained by \citet{shi14} using three dimensional shearing box simulations.  \citet{Young:2015frag} also show -- in the context of fragmentation (see below) -- that collapse of unstable modes is never initiated on scales $\ll H$.  These results imply that `gravito-turbulence' extends over a very small range of lengthscales. Additionally, there is no obvious indication of an energy cascade with dissipation occurring on the smallest scales. On the contrary, in global, thin disk simulations, \citet{CLC09} find that mode dissipation occurs through large scale, almost sonic shocks. In this respect, the dynamics introduced by local GI is not akin to what one usually calls turbulence. Morphologically, structures in shearing box simulations look more like turbulence than the high $m$, tightly wound spirals seen in global simulations. This might be an artifact of the shearing sheet/box approximation, but merits further investigation. Whether gravito-turbulence shares properties with isotropic, isothermal turbulence has important implications for fragmentation, as we discuss in Section \ref{sec-numerical}.

\paragraph{Large $H/r$} 
As the mass of the disk is increased, the disk becomes unstable at larger values of $H/r$ and higher values of $Q$, as predicted from the \cite{laubertin78} dispersion relation.  Once the disk-star mass ratio increases above $\approx 0.1$, the evolution of the instability changes character somewhat, transitioning from quasi-stationary spiral structures, to recurrent strong episodes of spiral activity, followed by brief quiescent phases. This behavior has been observed in a range of simulations including isolated disks with adiabatic equations of state with and without optically thin cooling \citep{LodRi05,mejia05,Laughlin:1997},  isothermal and irradiated embedded disks \citep{krumholz07,KMKK10}, and even two-dimensional, embedded disks with radiative heating and cooling \citep{Zhu:2012}.  A similar, recurrent behaviour has also been identified in $N$-body  galaxy simulations, where dynamical cooling occurs via the injection of low-velocity dispersion, or `cold' particles  \citep{SellCarl84}. 

The episodic behavior observed in global simulations is likely due to one of two phenomena. The first is the growth of the $m=1$ mode, as the disk mass becomes comparable to the stellar mass ($M_{\rm d} \geq 0.3M_*$, in analytic estimates \citet{ARS89}). Non-axisymmetric perturbations grow due to interaction with the``indirect potential" \citep{ARS89}. The indirect potential, caused by the displacement of the central body from the system center of mass, manifests as an extra term in the gravitational potential when working in a reference frame centered on the central mass. These modes  arise at higher values of $Q$ than the axisymmetric modes. Simulations of massive disks that neglect this term may underestimate the growth of instabilities. Under some circumstances, SLING amplification may occur \citep{Sling1990}, where the $m=1$ mode can amplify by interacting with the disk edge and the outer and inner Lindblad resonances. Because the $m=1$ mode leads to very large mass transport rates, the disk mass drains on such short timescales so that it can (at least temporarily) become stable and quiescent.   The second source of the quasi-periodic behavior is likely mode coupling, which we discuss in Section \ref{sec-modemode}.

\subsection{Accretion and transport}
\label{sec:accretion}
We now turn to the most important consequence of GI in disks: its ability to drive angular momentum transport, and therefore accretion of matter onto the central object. The fact that a spiral structure in the disk can transport angular momentum was recognized initially by \citet{lyndenbell72} in the context of galactic dynamics. The $R\phi$ component of the associated stress is given by
\begin{equation}
T_{r\phi}^{\rm grav} = \int\left\langle\frac{g_rg_{\phi}}{4\pi G}\right\rangle\de z,
\end{equation}
where $g_r$ and $g_{\phi}$ are the radial and azimuthal component of the perturbed self-gravity field and the brackets indicate azimuthal averaging. To this stress one should also add the induced Reynolds stress:
\begin{equation}
T_{r\phi}^{\rm Reyn} = \Sigma \langle\delta v_r\delta v_{\phi}\rangle,
\end{equation}
where $\delta v_r$ and $\delta v_{\phi}$ are the perturbed fluid velocities (see e.g. \citealt{balbus99}). For a magnetized disk (not considered here), one should also add the relevant Maxwell stress to the two terms described above. The  \citet{SS1973} $\alpha$ prescription relates the stress $T_{r\phi}$ to the local disk pressure by
\begin{equation}
T_{r\phi} = \alpha\Sigma c_{\rm s}^2\dln.
\label{eq:alpha}
\end{equation}
The dimensionless coefficient, $\alpha$, often called the anomalous viscosity parameter, is expected to be less than unity, To facilitate comparison with an eddy viscosity, this is often expressed as 
\begin{equation} \label{eq-alpha}
\nu = \alpha c_s H,
\end{equation}
implying that sub-sonic eddies on scales smaller than $H$ facilitate angular momentum exchange. 
The introduction of this parameterization is especially useful for broad parameter studies in a range disks at moderate computational cost by solving a 1D diffusion equation, often coupled with an energy equation, as we discuss in Section \ref{sec:alphaGI}. One can also  compute numerically the stress induced by GI (or another process like the MRI) by running a (magneto-)hydrodynamical simulation of an unstable disk, and compare the resulting total stress  to \Eq{eq:alpha} to obtain an effective $\alpha$. 

Let us now consider the torque and power associated with a spiral wave induced by GI. Standard wave mechanics \citep{toomre69,shu70,fanlou99} link the energy and angular momentum of a wave, $\mathcal{E}_{\rm w}$ and $\mathcal{L}_{\rm w}$ respectively, to the wave action $\mathcal{A}$:
\begin{equation}
    \mathcal{E}_{\mathrm{w}} = \omega \mathcal{A} = m \Omega_{\mathrm{p}}
    \mathcal{A}, 
     \label{wavedensities}
\end{equation}
\begin{equation}
    \mathcal{L}_{\mathrm{w}} = m \mathcal{A},
    \label{wavedensities2}
\end{equation}
where $\Omega_{\rm p}=\omega/m$ is the pattern speed of the perturbation. Thus the ratio of energy and angular momentum (and the ratio of power and torque exerted on a disk when launching the wave) is the pattern speed $\Omega_{\rm p}$, rather than the local $\Omega$ as for a viscous stress. More explicitly, the wave energy is given by 
\begin{equation}
  \mathcal{E}_{\mathrm{w}} = \frac{\Sigma}{2}\frac{m^{2}}{k^{2}}     \Omega_{\rm p}
    (\Omega_{\mathrm{p}} - \Omega) \left(\frac{\delta \Sigma}{\Sigma}
    \right)^{2},
  \label{waveenergydensity}
\end{equation}
where $\delta\Sigma/\Sigma$ is the fractional amplitude of the perturbation. \Eq{waveenergydensity} can be rewritten as the sum of two components:
\begin{equation}
  \begin{array}{lcl}
    \displaystyle
    \mathcal{E}_{\mathrm{w}} & = &\displaystyle
    \frac{\Sigma}{2}\frac{m^{2}}{k^{2}} (\Omega_{\mathrm{p}} - \Omega)^{2}
    \left(\frac{\delta \Sigma}{\Sigma} \right)^{2} \\     
    &  +  & \displaystyle \frac{\Sigma}{2}\frac{m^{2}}{k^{2}}     \Omega
    (\Omega_{\mathrm{p}} - \Omega) \left(\frac{\delta \Sigma}{\Sigma}
    \right)^{2}. \\       
    \label{energysplit}
  \end{array}
\end{equation}
The second term in the above expression is equal to the wave angular momentum times the local angular velocity and thus represents a local, viscous-like term. The first term is an `anomalous' energy transport term, which is inherently non-local \citep{balbus99}. The ratio of these two terms, $|\Omega_{\rm p}-\Omega|/\Omega$, is therefore a dimensionless parameter that measures the degree of non-locality of the disturbance. In practice, the transport induced by spiral waves will be global in character if the waves are able to travel within the disk significantly away from their corotation radius. 

We have seen already that the local dispersion relation involving the parameter $J$ (see \Eq{eq:j} above) predicts that the transition between local and global behavior is related to the mass ratio $M_{\rm d}/M_{\star}$. 
Is the mass ratio condition equivalent to a division of regimes based on the distance that waves propagate before dissipating? Clearly, to answer this question we have to look at the spectrum of modes excited by the instability. Using a spectral analysis of the dominant modes excited by GI, one finds a clear correspondence between morphology and the dominant mode of angular momentum transport. \cite{CLC09} computed the pattern speed of the dominant modes,  finding that independent of the disk-star mass ratio, the Doppler shifted Mach number $\mathcal{M}$ of the dominant perturbation is remarkably close to unity:
\begin{equation}
\mathcal{M}=\frac{|\Omega_{\rm p}-\Omega|r}{c_{\rm s}}\approx 1.
\end{equation}
This implies that the shocks that develop as soon as the waves travel far enough from corotation to become sonic quickly dissipate the waves. Based on this sonic condition it is now possible to translate the `locality' parameter defined after \Eq{energysplit}, into the disk-star mass ratio:
\begin{equation}
\frac{|\Omega_{\rm p}-\Omega|}{\Omega}\approx \frac{c_{\rm s}}{R\Omega}=\frac{H}{r}.
\end{equation}
For a $Q\approx1$ disk, $H/r\approx M_{\rm d}/M_{\star}$, thus the `corotation condition' and the `disk mass' condition are equivalent, so long as waves dissipate once they become sonic. We immediately see that the fraction of transport induced by non-local processes is proportional to the mass ratio.

\subsection{Saturation of the Instability vs Non-linear growth}
Thus far we have only discussed the linear growth of GI. However, to properly assess the outcome of the instability we need to consider how it evolves in the non-linear regime. We have introduced two dimensionless parameters, $Q$ and $M_{\rm d}/M_*$ that capture the linear phase. In order to understand the non-linear evolution, we require a third, independent dimensionless parameter, which captures the radiative timescale of the disk.

\subsubsection{Thermal Saturation}
The non-linear evolution of the instability and its saturation is best understood using numerical hydrodynamical simulations of self-gravitating disks, which have been the focus of most of the attention in this field over the last twenty years.
It is still useful to begin with a few simple analytic arguments.

Consider first an isolated disk with an initial surface density and temperature profile such that $Q \sim 1$ over a range of radii. In this scenario, $Q$ will evolve either due to mass redistribution or heating and cooling. We begin by examining changes in temperature rather than surface density because the viscous timescale is often slow compared to the thermal timescale. Since the excitation of waves will inevitably lead to shocks, it is natural to imagine a self-regulation mechanism (originally proposed by \citealt{pacinski78}) whereby the parameter $Q$ plays the role of an effective 'thermostat' for the disk temperature. In the absence of other significant heating terms, the disk cools down until it reaches marginal stability. At this point, GI sets in, causing shocks that heat the disk. If sufficient heat is retained, the disk will heat back up, quenching the instability, which in turn decreases the heating rate, allowing the disk to cool back towards instability. Thus in this self-regulated state, the disk should remain marginally unstable, with $Q$ close to unity. In such a state the disk permits growing modes, but the spiral perturbations do not grow exponentially, instead saturating at some finite value (see \Eq{eq:perturbationamplitude} below).

One can estimate analytically the conditions under which the disk should self-regulate. Since the instability provides heat on a timescale that is of order the dynamical timescale in the disk, $t_{\rm dyn}=\Omega^{-1}$, the `thermostat' can only work if cooling is slower than this timescale, that is if $\beta=\Omega t_{\rm cool}\gg 1$, where $t_{\rm cool}$ is the cooling timescale\footnote{Note that the requirement that the cooling timescale be shorter than the dynamical timescale in order to result in fragmentation has been discovered in a variety of contexts even outside the context of disk instability \cite{rees76,silk77,Thompson:1988}}. Thus we see that $\beta$ is a third, independent dimensionless number that controls not the onset of the instability like $Q$, or its morphology, like $M_d/M_*$, but its evolution and saturation.

The importance of the parameter $\beta$, and the relation to self-regulation, was first illustrated numerically in a paper by \cite{gammie01}. This work introduced an explicit cooling term in a self-gravitating shearing sheet calculation 
\begin{equation}
\left.\frac{\de u}{\de t}\right|_{\rm cool}=-\frac{u}{t_{\rm cool}},
\end{equation}
where $u$ is the internal energy of the fluid and $t_{\rm cool}$ is the cooling timescale, which is set as a free parameter in the simulations to be some constant times the dynamical timescale $\Omega^{-1}$. It is important to stress that the above description is not meant to reproduce any specific cooling law, but is a toy model for exploring the role of the cooling timescale in the outcome of GI. Different investigators use different prescriptions for the specific form of the cooling time, which in some cases is taken to be a constant \citep{mejia05} while in other cases is taken to be proportional to the dynamical timescale as a function of radius, so that $\beta=\Omega t_{\rm cool}$ is kept constant (e.g., \citealt{gammie01,LR04,LodRi05}). In particular, the so-called $\beta$-cooling prescription is now commonplace when simulating self-gravitating disks in a simplified manner. One noteworthy limitation of this  model is it can only capture the behavior of optically thin gas in numerical simulations, because it does not account for optical depth effects, which can be important. If one aims to capture more realistic thermal physics, radiative transfer must be properly accounted for \citep{Boley:2006, Stamatellos:2009, 2009SciKrum}. Nevertheless, the evolution produced by simplified cooling models is remarkably similar to that in more complex simulations.

In general, there is now a coherent picture emerging from all such simulations of isolated, non-irradiated discs, despite the large variety of hydrodynamic techniques (SPH or grid based codes, in 2D or in 3D) and cooling prescriptions employed. If cooling is relatively inefficient, so that the cooling timescale is long ($\beta\gtrsim 10-20$, though see Section \ref{sec-numerical}), the disk settles into a self-regulated state, where the spiral structure steadily transports angular momentum providing the heating required to balance cooling and keep the disk in thermal equilibrium. We call this thermal saturation of GI. We discuss in Section \ref{sec-numerical} uncertainties in the critical value of $\beta$ and 2D versus 3D effects.

As discussed above, if the disk is low mass (with $H/R\sim M_{\rm disk}/M\lesssim 0.1$), the transport induced by GI is local, and analogous to a viscosity.  If dissipation is the only heating source, then a disk undergoing GI in thermal equilibrium \citep{Pringle:1981,Gam2001}, will have the following relation between  $\alpha$ and the cooling rate:
\begin{equation}
\alpha=\dln^{-2}\frac{1}{\gamma(\gamma-1)\Omega t_{\rm cool}}.
\label{eq:alphasg}
\end{equation}
Since the energy density and the angular momentum density associated with density waves scales with the square of the perturbed gas density (e.g. Eq. \ref{waveenergydensity}), the condition for thermal regulation (Eq. \ref{eq:alphasg}) requires that the instability saturates at an amplitude that scales with the square root of the cooling time:
\begin{equation}
\frac{\delta\Sigma}{\Sigma}\propto \frac{1}{\sqrt{\beta}},
\label{eq:perturbationamplitude}
\end{equation}
a behavior confirmed by the numerical simulations of \citet{CLC09}. This relation is consistent with the critical $\beta$ being of order a few, since it is the point at which density perturbations have non-linear amplitude. In practice, energy dissipation occurs through the development of roughly sonic shocks.  This simple relationship suggests that the strength of the instability, and its efficiency as an angular momentum transport process, are correlated with the thermodynamics in steady state. We review GI models that use $\alpha$ as a proxy for transport in Section \ref{sec:alphaGI}.

\subsubsection{Thermal saturation in irradiated disks} 
\label{sec-irrad}
The discussion and the simulations described above assume that the only source of heating in the disk is that provided by the instability itself as it dissipates energy through shocks. However, protostellar disks are generally expected to be irradiated by either the central protostar, their natal envelope, or nearby stars. In the presence of irradiation, clearly the thermal saturation described above must change, since the disk alone is not responsible for setting the equilibrium temperature. The thermostat cannot function according to \Eq{eq:alphasg}.

When comparing GI in irradiated versus self-heated disks, one must consider the behavior at fixed $Q$, so that the disk has the same propensity to the instability. Even at fixed $Q$, the impact of irradiation is not intuitive. On one hand, in the presence of external heating a much lower level of dissipation due to the instability is required to prevent runaway growth, which would imply a stabilization of the disk. On the other hand, a very strongly irradiated disk will behave more like an isothermal disk, where the dissipation of energy due to even a large perturbation has little effect on thermal balance \citep{KMC11}. In this case it is hard to envision a mechanism that would stop an unstable perturbation from growing non-linearly, leading to fragmentation. Indeed irradiated disk simulations have more similarities to isothermal simulations than barotropic or cooling simulations \citep{2009SciKrum,Offner:2009}.  For disks in the moderately irradiated regime, both arguments apply. \citet{Rice:2011}  conducted local 2D simulations of disks including a `standard' $\beta$-cooling term and a constant heating term, representing the effects of irradiation. On the basis of cooling times,  irradiation stabilizes the disk in that a marginally faster cooling time is required for fragmentation in the irradiated case as compared to the non-irradiated case. However,  Eq. \ref{eq:perturbationamplitude} does not hold anymore, and for a given value of $\beta$ the average perturbation amplitude is much smaller. Thus, in the irradiated case, even perturbations with a very small average amplitude result in a runaway collapse into fragmentation. In this sense, irradiation has the effect of destabilizing the disk, because smaller amplitude perturbations can collapse. It is unclear whether the use of 2D simulations in this study might affect the measured critical $\beta$ \citep{Young:2015frag}.

In irradiated disks, there is another route to saturation that does not relate to thermal balance, but rather is dynamical, in the form of non-linear coupling between modes.

\subsubsection{Mode-mode coupling}
\label{sec-modemode}
The numerical investigation of self-gravitating disks began well before sophisticated radiative transfer cooling algorithms were computationally feasible. Early numerical simulations found that both isothermal \citep{laughlin94}, and barotropic \citep{laughlin96,laughlin98} disks could develop strong gravitational instabilities that lead to fast redistribution of angular momentum within the disk. In these regimes, GI cannot self-regulate in the manner describe above as they represent $\beta\rightarrow0$ and $\beta\rightarrow\infty$ respectively. Instead,  non-linear mode coupling appears responsible for saturating the amplitude of GI-induced perturbations. \citet{Laughlin:1997} demonstrate that coupling between $m=0$ and $m=2$ modes prevents  runaway growth and thus fragmentation, while allowing rapid angular momentum transport. This same behavior is evident in the isothermal simulations of \citet{KMKK10}, and is also likely responsible for the transient spiral patterns seen in the massive disk simulations of \citet{LodRi05}. 

The exact conditions under which mode-mode coupling dominates remain unclear, as little detailed work has been done since \citet{Laughlin:1997}. Based on the simulations in which a similar effect has been observed, we expect this saturation mechanism to dominate in massive disks, as well as disks that are irradiation dominated. These two cases represent (1) the regime in which local dissipation breaks down and (2) the regime in which the energy balance due to GI induced heating and radiative cooling breaks down. Ironically, this saturation mode may be the dominant mechanism preventing fragmentation in protostellar disks, despite the dearth of attention in the literature.

\subsection{When saturation fails: Fragmentation} 
\label{sec-satfail}
If the saturation mechanisms listed above fail to prevent runaway growth of perturbations, the unstable region of the disk will break up into bound objects, or fragments. As the perturbations grow to non-linear amplitudes, their own self-gravity causes them to collapse and separate from the background disk. Fragmentation is in some ways an alternative means for shutting off the instability: the separation of large amplitude perturbations form the disk means a reduction in local surface density, and thus an increase in $Q$.
In the context of protostellar disks this is of great importance, as fragmentation can in principle lead to the formation of either stellar, brown dwarf, of perhaps planetary mass companions. We provide a brief overview of the two common pathways to fragmentation, deferring a discussion of the astrophysical implication to Sections \ref{sec-sgprotostellar} and \ref{sec-sgplanet}

\paragraph{Fragmentation driven by cooling}
A great majority of disk simulations have considered disk fragmentation as consequence of rapid cooling in isolated disks. In this case, $Q$ decreases due to a progressive drop in temperature, until GI sets in. If the cooling is rapid, GI cannot provide sufficient heating to counteract cooling, and the disk will fragment. 
Exactly how fast the disk must cool (the so-called critical value of $\beta$)  has been the subject of many investigations. Studies have examined the role of the temperature dependence of cooling \citep{Johnson:2003,CLC10}, the role of the evolution of $\beta$ in time at fixed radius \citep{CHL07}, and the role of irradiation \citep{Boley:2007,Stamatellos:2008, Rice:2011}. The critical value is a function of the equation of state, and is of order $\beta=10-20$. The deviations from the critical value of $\beta$ due to more complex physical effects are of order a factor of two. We review the numerical complications of measuring this value in Section \ref{sec-numerical}.

Cooling as a path to fragmentation is most likely relevant for disks that are thermally self-regulated because, as discussed above, irradiation and global transport can hamper the operation of the disk thermostat. AGN disks, or protostellar / protoplanetary disk which are shadowed by inner disk wall or warps may be the best candidates for this type of fragmentation. Alternatively disks which undergo rapid changes in accretion rate, which in turn can effect the stellar luminosity and thus the irradiated disk temperature, might also be subject to fragmentation driven by cooling \citep{KMC11}.

\paragraph{Fragmentation driven by accretion}\label{sec-sgphys-infall}
The second path to fragmentation is due to evolution of the disk surface density either due to infall from the external environment, or due to changes in the accretion rate within the disk that lead to mass pile-ups. For disks undergoing infall at some rate $\dot{M}_{\rm in}$, there exists a regulation mechanism akin to thermal saturation. Consider a disk with $Q>1$ that accretes material such that $Q$ declines towards unity. As the disk approaches $Q\approx1$, the linear instability transports angular momentum. If the accretion induced by GI gives $\Mdot<\Mdot_{\rm in}$, the disk will become more massive, and thus susceptible to a wider variety of modes, enabling more transport. If $\Mdot_{\rm in } \leq \Mdot_{\rm max,GI}$, the maximum possible rate that GI can provide (as yet unspecified, see below), the disk will transport material at the rate it is being fed, with the disk mass acting as the regulator of transport \citep{KMKK10, Zhu:2012}. If $\Mdot_{\rm in} > \Mdot_{\rm max,GI}$ the disk will ultimately fragment \citep{KMK08}.  The ability of the disk to regulate transport up to a critical rate is likely related to the spiral-mode coupling described above.

What sets the maximum transport rate induced by GI, and how does one know when disks will fragment versus increase in mass? \cite{KMKK10} argued that two dimensionless parameters, benchmarked to the infall rate, are strong predictors of fragmentation driven by accretion. These two are 
\begin{eqnarray}
\xi = \Mdot_{\rm in}/(c_{\rm{s}}^3/G)\\
\Gamma = \Mdot_{\rm in} /(M_{*d}\Omega), 
\end{eqnarray} 
where $M_{*d}$, the total system mass. The first parameter, $\xi$ references the infall to the isothermal sphere collapse rate, $c_s^3/G$ \citep{Shu77} but uses the disk, not core temperature. For a disk in steady state, $\xi$ and $\alpha$ are interchangeable, using the steady-state accretion relation $\Mdot = 3 \alpha c_s^3/(GQ)$ \citep{Frank:2002}. However in many cases, the disk accretion rate does not come into equilibrium with the infall rate. The second parameter, $\Gamma$ is the ratio of the orbital timescale to the disk mass doubling timescale. When infall is significant, $\Omega$ is controlled by the circularization radius of accreting material (not viscous spreading) and is thus a proxy for disk angular momentum.

 Using a suite of 3D numerical simulations of isothermal disks undergoing self-similar accretion, \citet{KMKK10} find that disks undergoing infall with $\Gamma <\xi^{2.5}/850$ are unstable to fragmentation.
Subsequent work by \cite{OKMKK10} and \cite{Zhu:2012} confirm these relations hold when more complicated thermal physics and more realistic, turbulent infall conditions are included. Such a parameterization is applicable to AGN disks as well, where it might be relevant for seed black hole growth at high redshift \citep{LN06}.

The $\xi$, $\Gamma$  parameterization predicts that the maximum transport rate is a function of mass ratio (in steady state with $Q\sim1$, we have that $H/R\approx M_{\rm d}/M_* \approx (\Gamma/\xi)^{1/3}$). For low mass disks, where GI-induced transport is local, \citet{RLA05} have shown that the existence of a critical cooling rate for fragmentation translates, through  \Eq{eq:alphasg}, into a more general upper limit of $\alpha \approx 0.1$. The boundary for fragmentation in $\xi-\Gamma$ space is consistent with this value of $\alpha$ for equivalently low mass disks, despite the absence of thermal regulation. If the disk mass grows above $0.1-0.2M_*$, however, transport becomes inherently global, and thus an increase in mass ratio corresponds to an increase in the maximum rate of transport. This is consistent with the \cite{laubertin78} dispersion relation. According to the $\xi-\Gamma$ boundary derived above, a more massive disk corresponds to a larger value of $\Gamma$ for fixed $\xi$, which will thus be stable at the same infall rate. These simulations suggest that global modes can provide close to $\alpha \approx 1$ when  $M_{\rm d} \rightarrow M_*$. 

\paragraph{Other fragmentation triggers}
 \citet{nelson2000,mayer05b,boss06} and \cite{LMCR07} have studied the influence of a massive companion on disk fragmentation, reaching apparently contradictory conclusions. While \citet{boss06} argues that the presence of a companion increases the tendency for fragmentation, all of the other studies obtain the opposite result. They find that the tidal heating due to the companion has the effect of inhibiting fragmentation in a marginally stable disk. 
 
\subsection{Capturing GI with $\alpha$ models}\label{sec:alphaGI}
As mentioned in the previous sections, it is often convenient to parametrize angular momentum transport measured in numerical simulations in terms of the \citet{SS1973} $\alpha$-parameter. We have relied on this parameterization heavily in discussing thermal saturation (see \Eq{eq:alphasg}). Despite the inherent assumption of local transport in the $\alpha$-model \citep{balbus99}, the approximation is also often used for global transport, where it approximates the accretion rates measured remarkably well. \cite{KMK08} compiled transport measurements from numerous simulations, and found that a two component $\alpha$ model -- one to account for local transport, one for global, could accommodate a wide range of disks in terms of both mass and thermal physics.
As noted above, the biggest advantage of invoking an effective $\alpha$ to capture GI is that one can conduct parameter studies over a wide swath of physical variables using semi-analytic or 1D models. We provide a brief overview of the most common models used.

The first proposed $\alpha_{\rm sg}$ models of \citet{linpringle87,linpringle90} posit that since $Q$ controls the onset of instability, the strength of transport should scale inversely with $Q$. They propose:
\begin{equation}
\alpha_{\rm sg}=\left\{
\begin{array}{cc}
\zeta\left(\displaystyle\frac{\bar{Q}^2}{Q^2}-1\right)  & \hspace{2cm} Q<\bar{Q}  \\
\\
0                                                                                        & \hspace{2cm} Q>\bar{Q} 
\end{array}
\right.
\label{eq:linpringle}
\end{equation}
where $\zeta$ is an additional parameter measuring the strength of the gravitationally induced torque, and $\bar{Q}$ is the maximum value ($\approx 2$) at which GI sets in. This prescription, at face value, is orthogonal to  $\alpha$ models based on \Eq{eq:alphasg}. 
Although the dependencies appear distinct between  \Eq{eq:alphasg} and \Eq{eq:linpringle}, they can be tuned to capture similar behavior, but are better suited to different regimes. \Eq{eq:alphasg} is appropriate for disks that are in steady-state and thermally self-regulated. It cannot be used to capture the behavior of a disk with $Q$ much different from unity, or one that is out of equilibrium because the relation is predicated on energy balance. On the other hand, \Eq{eq:linpringle}  can capture GI in a disk that is non-self regulated and may be evolving (either in surface density or temperature) in time. Because \Eq{eq:linpringle} has a free normalization parameter, it can also describe a disk in steady-state that is marginally unstable by tuning $\zeta$ so that the transport rate matches up with the R.H.S. of \Eq{eq:alphasg}. 


These  $\alpha$ models have proved indispensable for studying multiple transport processes simultaneously. \cite{Armitage:2001} explored 1D viscous models including both MRI and GI-induced transport, showing  time variable accretion capable of reproducing FU-Ori like episodic accretion. \citet{BL2001} discuss how GI would make the outer disk in FU Ori objects hotter compared to non self-gravitating disks and hence produce a flatter SED (see also \citealt{ARS89}). \cite{Martin:2011} similarly identify regions of disk parameter space lacking steady-state solutions as responsible for outbursting behavior, in a fashion akin to the limit-cycle instability for dwarf novae. \citet{ML2005} first identified, through simple $\alpha$ models, the existence of a critical radius at $\approx 50-100$ AU beyond which the disk is expected to fragment. \cite{Clarke:2009} use an $\alpha$ model for GI to estimate the relevant cooling time as a function of radius, and show that fragmentation occurs beyond $\sim 70$ AU, while self-regulation is expected between $\sim 20-70$ AU. Crucially, \citet{Clarke:2009} neglected the effect of irradiation from the central star, that in general makes the inner disk gravitationally stable rather than self-regulated.

Recently, \cite{Rafikov:2015}, following \cite{Rafikov:2009}, developed a more comprehensive viscous $\alpha$ prescription, which encompasses a wider range of disk states, although the caveats about locality of energy deposition remain. \cite{Rafikov:2015} finds that the predicted disk properties are quite similar to those derived under  $\alpha(Q)$ models (e.g. \citet{KMK08,Zhu:2010}). The important feature of these viscous models is the smooth growth of transport rates as $Q$ declines, with a cutoff at some maximum $\alpha$ for a minimum $Q$.

\subsection{Convergence of numerical results}\label{sec-numerical}
In our analysis of the saturation of GI, we have relied heavily on numerical simulations. The two values often pulled from these simulations are the critical value of $\beta$, or the maximum transport rate $\alpha$, at which saturation fails, leading to fragmentation. These two quantities (directly correlated in some regimes) are only measurable in simulations. Frustratingly, these quantities can be difficult to disentangle from numerical artifacts. No astrophysical simulations operate at realistic Reynolds numbers,  and all suffer from numerical errors that manifest as an effective viscosity. In the case of SPH simulations, an artificial viscosity must be included explicitly.  For grid codes, \cite{krumholz07} for example, has measured the effective $\alpha$ from diffusion as a function of resolution, showing that it can be competitive with physical mechanisms at moderate resolution.

At the time of this writing, the convergence problem in self-gravitating disks simulations is an open issue. We review them below for completeness, but emphasize that for application to protostellar and protoplanetary disks, uncertainty in the critical value of $\beta$ has little impact on disk evolution. Indeed, in the region where protostellar disks are gravitationally unstable, the cooling time is a steep function of disk radius, so that changes in the critical value of $\beta$ only lead to minor changes to the radius at which fragmentation occurs \citep{Clarke:2009,CL09}.

Until recently, most investigations of resolution focused on resolving fragmentation. Early studies found that the  Jeans length \citep{bate97,Truelove97} must be resolved by $\approx 10$ grid cells to accurately capture fragmentation. More recent work has argued that the Jeans length must be resolved by as many as 64 grid cells to achieve convergence \citep{Turk:2012}. Since in SPH the resolution is set by the smoothing length $h$, and the Jeans length is of the order of the disk thickness $H$, the Truelove and Bate \& Burkert criteria imply that $h/H\lesssim 1$ to resolve fragmentation. This condition is amply satisfied by most recent simulations. \citet{nelson06} lists three different conditions: one is a variant of the Jeans length criterion, the second is that $h/H\lesssim 1$, and the third is that adaptive smoothing lengths be used in SPH (which is done in all modern SPH calculations). That convergence remains a challenge even when the above resolution criteria are met is not so surprising, since none explicitly capture one of the most important aspects of the physics: dissipation.

\citet{meru11a,meru11b} found that  in simulations with several million SPH particles (factors of several above most previous work) the fragmentation criteria as measured by a critical $\beta$ did not appear to  converge. The disk fragmented at larger $\beta$ with higher resolution. Similar results have come from grid-based simulations. \cite{Paardekooper:2011} found that fragmentation could be artificially triggered at sharp boundaries between the stable and unstable portions of the disk, while \cite{Paardekooper:2012} claimed, based on 2D simulations, that fragmentation might be a stochastic process, and therefore allowed  (rarely) at values of $\beta$ associated with much smaller amplitude perturbations. This interpretation would be consistent with gravito-turbulence fully sampling a turbulent power spectrum of density perturbations \citep{Hopkins:2013}. These results cast doubt on the numerical value of the fragmentation boundary obtained in previous investigations (that pointed to a critical value $\beta_{\rm crit}\approx 10$). 

\citet{LodatoClarke11} address the importance of the magnitude of the artificial viscosity as a function of resolution, which inherently alters the disk thermostat. In SPH codes, artificial viscosity provides an equivalent $\alpha_{\rm art}\propto h/H$.  \citet{LodatoClarke11} find that in order for artificial viscosity to be negligible as a source of heating, one requires
\begin{equation}
\frac{h}{H}\ll \frac{40}{\beta}.
\end{equation}
Note that as $\beta$ is increased (as one does when probing the fragmentation boundary), the resolution requirement becomes progressively harder to satisfy, because a smaller amount of numerical dissipation represents a larger fractional contribution to heating.
Such a trend of increasing resolution requirements with increasing $\beta$ is indeed consistent with the \citet{meru11a,meru11b} results, but do not fully explain them, as it turns out that, based on these simple arguments, fragmentation would be artificially quenched by numerical viscosity when the latter produces only 5\% of the heat required by thermal equilibrium. 
\citet{Meru:2012}, on the other hand, point out that most SPH simulations for self-gravitating disks have employed an artificial viscosity formulation that strongly reduces viscosity for high Mach number shocks. Paradoxically, this reduction of the viscosity coefficient has the counter-intuitive effect of increasing artificial dissipation, since it generates artificially large velocity gradients \citep{LP10}.

Even more extreme challenges to the fragmentation boundary have come from \cite{Paardekooper:2012} and \cite{Hopkins:2013} who suggest that fragmentation may be inherently stochastic, and thus allowed at much longer cooling times. This relies on GI  producing a wide power spectrum of turbulent fluctuations
 Investigations by \cite{Young:2015frag} question this for two reasons. First, they show  that independent of numerical method, there are inherent limitations in the treatment of the small scale gravitational field in 2D. As a result, measurement of the fragmentation boundary (or stochastic fragmentation) in a 2D simulation is questionable, and better left to 3D simulations. They further argue that because of the quasi-regular nature of self-gravitating structures, stochastic fragmentation should be inhibited at very large $\beta$, because the growth and development of spiral structure occurs over a modest number of orbital periods, and thus may not be akin to a true turbulent cascade (see also \citet{shi14}).
 
Despite the lingering questions about resolution and critical cooling times, we emphasize that the critical $\beta$ for fragmentation has very little impact on whether disk fragmentation, and especially planet formation via GI, is viable because most disks that have $Q\sim1$ also attain relatively small values of $\beta$ for realistic protostellar disk conditions.

\section{Application to Protostellar Disks}\label{sec-sgprotostellar}
In Section \ref{sec-sgphys} we have given an overview of the basic physics that governs the behavior of disks when self-gravity becomes important. These principles apply to disks in any astrophysical context. Here we narrow our focus to the case of protostellar and protoplanetary disks in order to understand which of the regimes and corresponding behaviors are most relevant. The details of how star formation proceeds inevitably effects the characteristics of protostellar disks, and thus the importance of self-gravity and GI. It is beyond the scope of this review to describe how different modes of star formation might produce different disk properties; instead we endevaor to provide a broad overview of all plausible disk states.

\paragraph{Narrowing the Parameter Space}
We have shown that GI can be described by three independent dimensionless numbers that describe the state of the disk, $Q$, $M_{\rm d}/M_*$ and $\beta$. However, a protostellar or protoplanetary disk does not attain arbitrary combinations of these three quantities because the disk's internal energy and cooling rate are both temperature dependent. Consider $\beta$ in terms of these physical variables: 
\begin{equation}
\label{eq:tcooldisk}
t_{\rm cool} = \beta \Omega^{-1} = \frac{4}{9\gamma(\gamma-1)}\frac{\Sigma c_{\rm s}^2}{\sigma T^4}\tau
\end{equation}
where $\tau=\kappa\Sigma/2$, and $\kappa$ is the opacity. 
The cooling time, and thus $\beta$, depend on the disk surface density, temperature, and dust opacity. Once one selects a disk model with a given $Q$ profile and mass ratio, the profile $\beta(r)$ is also determined (albeit with uncertainty due to dust opacity models). We present typical disk scalings for temperature, mass, size, and accretion rate in order to survey GI in protostellar and protoplanetary disks, highlighting three case studies.

\subsection{Setting the Thermal Physics}
We begin with a simple model for the temperature profile of the disk. Ideally disk temperatures should be found from accurate radiative transfer calculations, which account for wavelength dependent opacities and heating sources. Nevertheless, the agreement between the simple analytic formulae below and radiative transfer calculations is quite good, making these useful for estimation purposes \citep{Zhu:2012,Boley:2010}.

Disks are heated primarily by three sources: accretion energy, host star irradiation, and external radiation. The disk midplane temperature can be found via the following relation by balancing heating and cooling terms:
\begin{equation}\label{eq-disktemp}
\sigma T^4 = \frac{3}{8}f(\tau) F_{\rm acc} + F_*+ F_{\rm ext}
\end{equation}
where $F_*$ and $F_{\rm ext}$ are the energy fluxes associated with stellar irradiation and external heating respectively, while $F_{\rm acc}$ is the energy flux associated with accretion and 
\begin{equation}\label{eq-ftau}
f(\tau)=\tau +\frac{1}{\tau }
\end{equation}
\citep{Rafikov:2005}. Here, $\tau$ is the Rosseland mean opacity, and $f(\tau)$ reasonably captures how the accretion energy  diffuses outward from the midplane in both the optically thick and thin regimes. We now examine each term in detail.

The first term on the R.H.S, $F_{\rm acc}$, describes the release of gravitational energy as material moves through the disk, falling deeper into the potential well. This is commonly referred to as viscous dissipation.  Independent of whether the disk is truly viscous, the potential energy of accreted material must be released, some portion of which will go into the thermal energy of the disk. The flux (far from the stellar surface) associated with this term if all of the accretion energy goes into heating is:
\begin{equation}\label{eq-facc}
F_{\rm acc} = \frac{3}{8\pi}{\Mdot\Omega^2}
\end{equation}
Because the heat must propogate from the midplane to the surface, this term is modified by the optical depth. As discussed in Section \ref{sec-obsev}, disk opacities are uncertain, and likely evolve in time due to grain growth and changes in the dust-to-gas ratio. At very early times, when grains are ISM-like ($\mu m-10\mu m$), the opacity is temperature dependent, typically $\kappa_R\approx \kappa_2 T^2$ (where $T$ is in Kelvin), \citep{Bell:1994}, in the cold, icy limit $T<155$K relevant for GI. \cite{Semenov:2003} find that the Rosseland mean opacity for small grains is well fit by $\kappa_2 \approx 5 \times 10^{-4} \rm{cm}^2/\rm{g}$. Once grain growth commences, grains quickly become large compared to the blackbody wavelengths of the disk. In this regime, the opacity is both lower and temperature independent. Based on  \cite{Pollack85}  we use a temperature independent $\kappa_0 = 0.24 \rm{cm}^2/\rm{g}$. We benchmark our results to this large grain case for several reasons. First, after no more than $10^5$yrs, there is good evidence that grain growth has commenced.  Although GI may be present earlier, at this epoch the disk is optically thick for either opacity law. Secondly, lower opacities associated with grain growth produce a wider range of disks susceptible to GI; since our intent is to review all the possible phase space we consider this choice ``conservative."

The second term on the R.H.S. of \Eq{eq-disktemp} captures irradiation from the host star. The stellar luminosity comprises two components: accretion luminosity and the intrinsic stellar luminosity that comes either from gravitational radiation or nuclear burning depending on age and mass. For low mass stars the accretion luminosity typically dominates. High mass stars, which reach the ZAMS while accreting, are dominating by H-burning \citep{Tout}. Unlike the dissipation of accretion energy, the impact of stellar irradiation on disk temperatures is highly dependent on the envelope and disk structure, because the absorption of radiation depends on both the intervening material reprocessing the radiation and on the flaring angle of the disk absorbing radiation. We consider two models for stellar irradiation. The first, from ray tracing calculations of \cite{ML2005}, finds that for deeply embedded disks, the heating term is well described by:
\begin{equation}\label{eq-mlirrad}
F_{*,{\rm emb}}  = \sigma T_{*,{\rm emb}}^4= f \frac{L_*}{4\pi r^2}
\end{equation}
with $f\approx 0.1$ derived from ray tracing calculations where much of the stellar flux impacts the infalling envelope (ala \citet{Terebey:1984}) and is reprocessed back down onto the disk. Note that $L_*$ can include both internal luminosity and stellar accretion luminosity. The second model refers to the case of a non-embedded disk, for which the analytic model of \cite{Chiang:1997} accounts self-consistently for the flaring due to incident irradiation, and heating of dust grains in the outer disk layers. They find:
\begin{equation}
F_{*,{\rm iso}} =  \sigma T_{*,{\rm iso}}^4 = \eqfrac{\alpha_F}{4}\eqfrac{R_*}{r}^{2} \sigma T_*^4
\end{equation}
where $\alpha_F$ measures the grazing angle at which starlight hits the disk; this in turn is a function of the height of the disk photosphere, which depends on grain properties. When disks reach the threshold for instability, they are typically optically thick, making the use of this approximation reasonable. We can rewrite this explicitly as a function of stellar luminosity and disk radius as:
\begin{equation}
T_{*,\rm iso} = \left[\frac{1}{(7 \pi \sigma)^2}\frac{k_b}{\mu}\frac{1}{GM_*}\right]^{1/7}L_*^{2/7}r^{-3/7}
\end{equation}
We note that when using this relation for temperature due to stellar irradiation in conjunction with other heating terms, there is a small inconsistency in that the actual disk scale height will differ slightly from that solved for in the irradiation model. 

Including both heating terms, we see that an individual disk may span both the self-luminous and irradiated regime at different radii. It is instructive to examine the radial scaling of the first and second terms in \Eq{eq-disktemp}:
\begin{eqnarray}
F_{\rm acc}  \propto \Mdot(r) r^{-3} \\
F_{*,{\rm emb}}  \propto L_* r^{-2}\\
F_{*,{\rm iso}}  \propto L_*^{8/7}r^{-12/7}
\end{eqnarray}
where the last two lines cover the embedded and isolated cases. In either case, the viscous term will fall off more steeply with radius than either irradiation term. Thus the inner disk is more likely to be self-luminous, while outer disk may be dominated by irradiation from either the central star or even the external environment. 

Finally, external radiation from other stars can contribute to the disk temperature (see \citealt{Thompson:2013} for an extreme example). We consider the interstellar radiation field to create a temperature bath of $T_{\rm ext}$ (such that $F_{\rm ext}=\sigma T_{\rm ext}^4$), which sets a lower threshold for the disk temperature as described above. In more active, higher mass star forming regions this can be up to $T_{\rm ext} = 20-30K$ \citep{Caselli:1995}. For deeply embedded cores, the envelope may insulate the disk from external sources, providing a somewhat lower temperature thermal bath. Neither irradiation term is accompanied by an optical depth because they set the temperature at the surface, which controls the cooling (L.H.S. of \Eq{eq-disktemp}).

\subsection{Disk Masses and Radii}
In Section \ref{sec-obsev} we showed evidence for disks ranging from $M_d/M_* \approx 0.001-0.1$, but argued that in many cases these may be lower limits based on uncertain grain growth,  dust-gas conversions, and biases against observing the youngest, most massive disks. Here we consider disks with $0.05<M_d/M_*<0.5$ to encompass both the least massive disks where self-gravity can be relevant, and a generous upper limit on those observed. Recalling \Eq{eq-mdhr}, one can immediately see that a $Q=1$ disk at the upper boundary has an aspect ratio so large as to call into question the disk geometry.

Direct observations show that disks extend out to $50-500$au, based on submm and optical absorption of proplyds \citep{Andrews:2009,Eisner:2008}. Disk radii at early times (Class 0, I phases) are controlled by the infalling angular momentum from the envelope or feeding filament \citep{Terebey:1984}. Most of the disk mass will be contained within the circularization radius:
\begin{equation}
r_d = R_c = \frac{\langle j \rangle^2}{\sqrt{GM}} 
\end{equation}
where $\langle j \rangle $ is the average specific angular momentum of accreting material, which can be obtained from considering
\begin{equation}
\dot{\bf{J}}(t) =  \Mdot(t)\bf{j}(t).
\end{equation}
In a simple core-collapse model one can specify $\bf{j}(t)\def\bf{j{(r}}$, however it is more realistic to consider the impact of accretion from turbulent filamentary structures that characterize star forming regions \citep{OKMKK10}. Turbulence in the interstellar medium follows a line-width size relation, such that increasingly larger scales will typically carry more angular momentum \citep{Larson:1981}. \cite{KM06} used this to generate models for disk radii as a function of time. It is not guaranteed, however, that the direction of the angular momentum vector will remain constant throughout infall, and thus disks may not  grow monotonically in time. Recent simulations by \citet{BLP10} and \cite{Fielding:2015} show that there can be enough change in the infalling angular momentum vector to tilt the disk with respect to the spin-axis of the star, especially if variable accretion causes substantial oscillations in the disk mass, and if stellar interactions truncate the disk. 

When disks transition from the protostellar to protoplanetary phases, infall of mass and angular momentum becomes less important. In this case one expects the disk to evolve towards the self-similar viscous spreading solution of \cite{lyndenbell74}. While this will lead to larger outer disk radii than that given by the circularization radius, the outer regions typically hold very little mass and thus do not have a significant impact on disk self-gravity.

In order to determine $\Sigma(r)$, we require a model for the radial profile. Here again we rely on observations, which suggest a power law, or a power law plus exponential tail \citep{Andrews:2009, Perez:2012}. Since the exponential part contains an irrelevant amount of the mass, we consider only the power law component. For our calculations, we choose  $\Sigma\propto r^{-1}$, which puts more mass at large radii where instabilities might arise, lest we underestimate the role of self-gravity. Note that this surface density profile, coupled with an irradiation dominated temperature profile is consistent with steady-state disk models with $\Mdot(r)=$const.
To give a quantitative idea of the scaling for our parameter ranges:
\begin{eqnarray}\label{eq-sigexamp}
\Sigma(r) =  \Sigma_0 \left(\frac{r}{R_{\rm out}}\right)^{-1}\left\{\begin{array}{cl}
\displaystyle \Sigma_0= 28{\rm g/cm^{-2}}\;\;& , \frac{M_d}{M_*} = 0.05,R_{\rm out} = 50{\rm au} \\
\rule{0ex}{5ex} \displaystyle \Sigma_0 = 2.8{\rm g/cm^{-2}} &,\frac{M_d}{M_*} = 0.5,R_{\rm out} = 500{\rm au} \\
\end{array}\right.
\end{eqnarray} 

\subsection{Accretion Rates}
The only remaining parameter to specify is the accretion rate through the disk, which has two effects. First, as discussed in Section \ref{sec-sgphys-infall}, infall of material on to the disk impacts the disk mass, angular momentum transport rate, and  susceptibility to fragmentation. Secondly, the disk temperature is a function of the accretion rate. We consider disks with steady-state accretion at rates ranging from $\Mdot = 10^{-8}-10^{-3} \Msun yr^{-1}$, where the lower end is consistent with accretion rates observed on actively accreting T Tauri stars towards the end of their lives, and the highest rates represent a generous upper limit set by infall from a supersonic core for more massive stars \citep{MT2003,Klaassen:2011} .

In reality, a disk with a given surface density and temperature profile cannot support an arbitrary accretion rate in steady state. A physical mechanism for providing angular momentum transport is required.  Indeed there is no guarantee that a disk will typically be in steady-state with $\Mdot(r)= $ const \citep{Armitage:2001,Martin:2011}.

As discussed in Section \ref{sec-sgphys}, saturated GIs provide transport at rates such that $10^{-2} <\alpha< 1$ depending on the disk parameters. Obviously magnetic effects, both small scale turbulence from the MRI \citep{BH94,Simon:2013}, or large scale magnetic winds \citep{Bai:2015} may also be important for various disk properties at different times. 
\begin{figure}[t]
\begin{minipage}{.62\textwidth}
\centering
\hspace{-1.1in}
\includegraphics[width=3.3in,left]{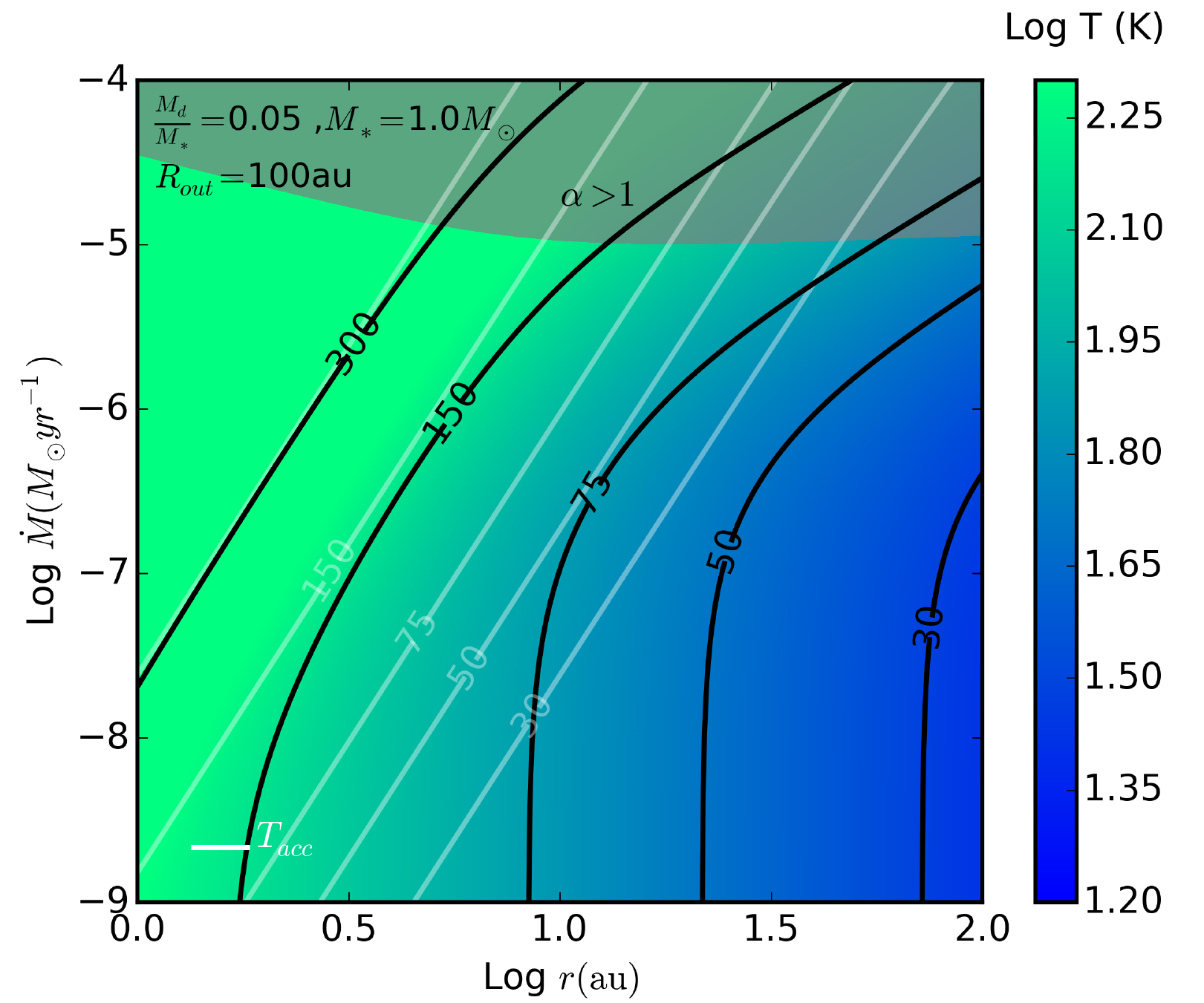}
\end{minipage}%
\begin{minipage}{.62\textwidth}
\centering
\hspace{-0.8in}
\includegraphics[width=3.3in,left]{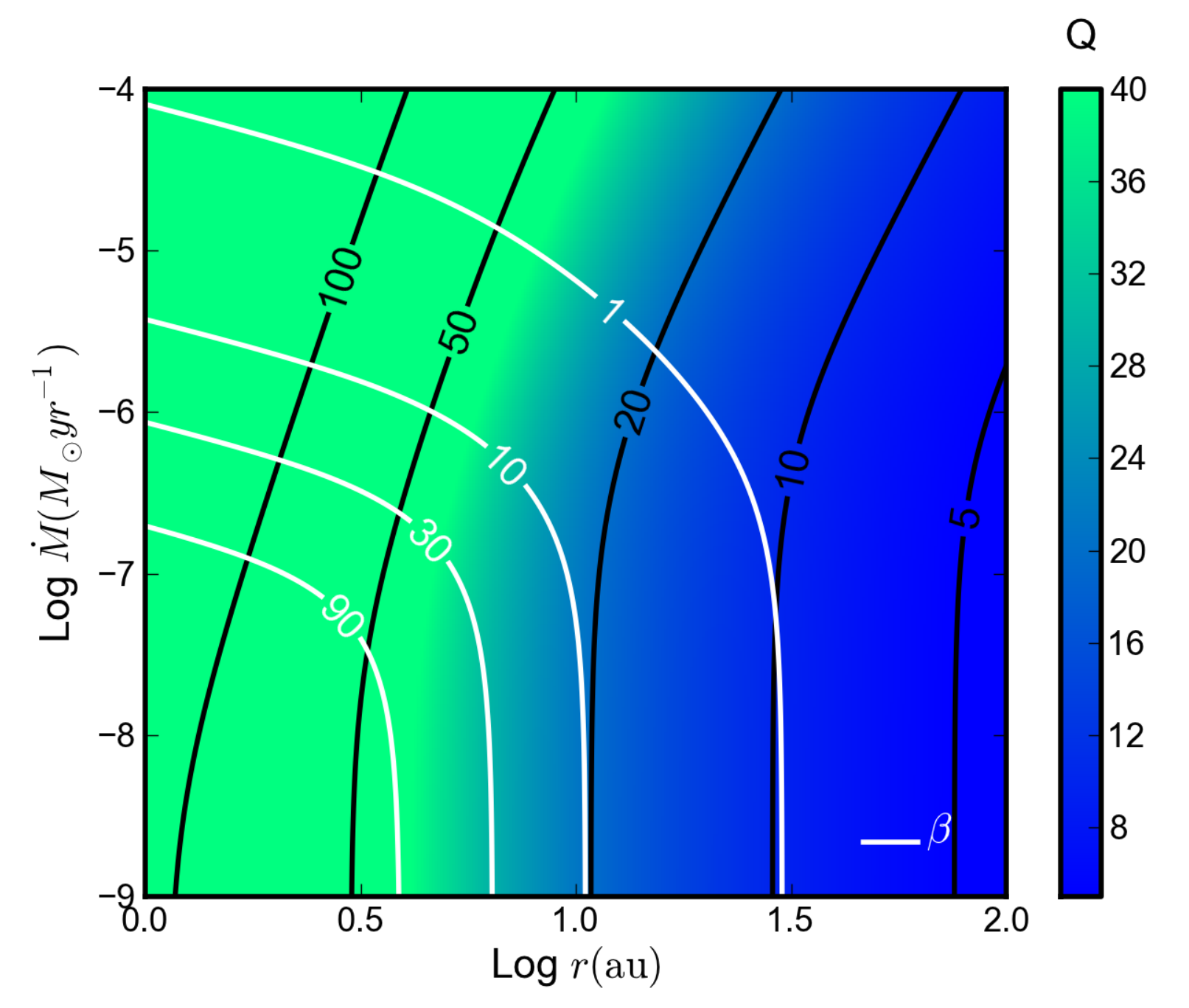}
\end{minipage}%
\end{figure}
\begin{figure}
\begin{minipage}{.62\textwidth}
\centering
\hspace{-1.1in}
\includegraphics[width=3.3in,left]{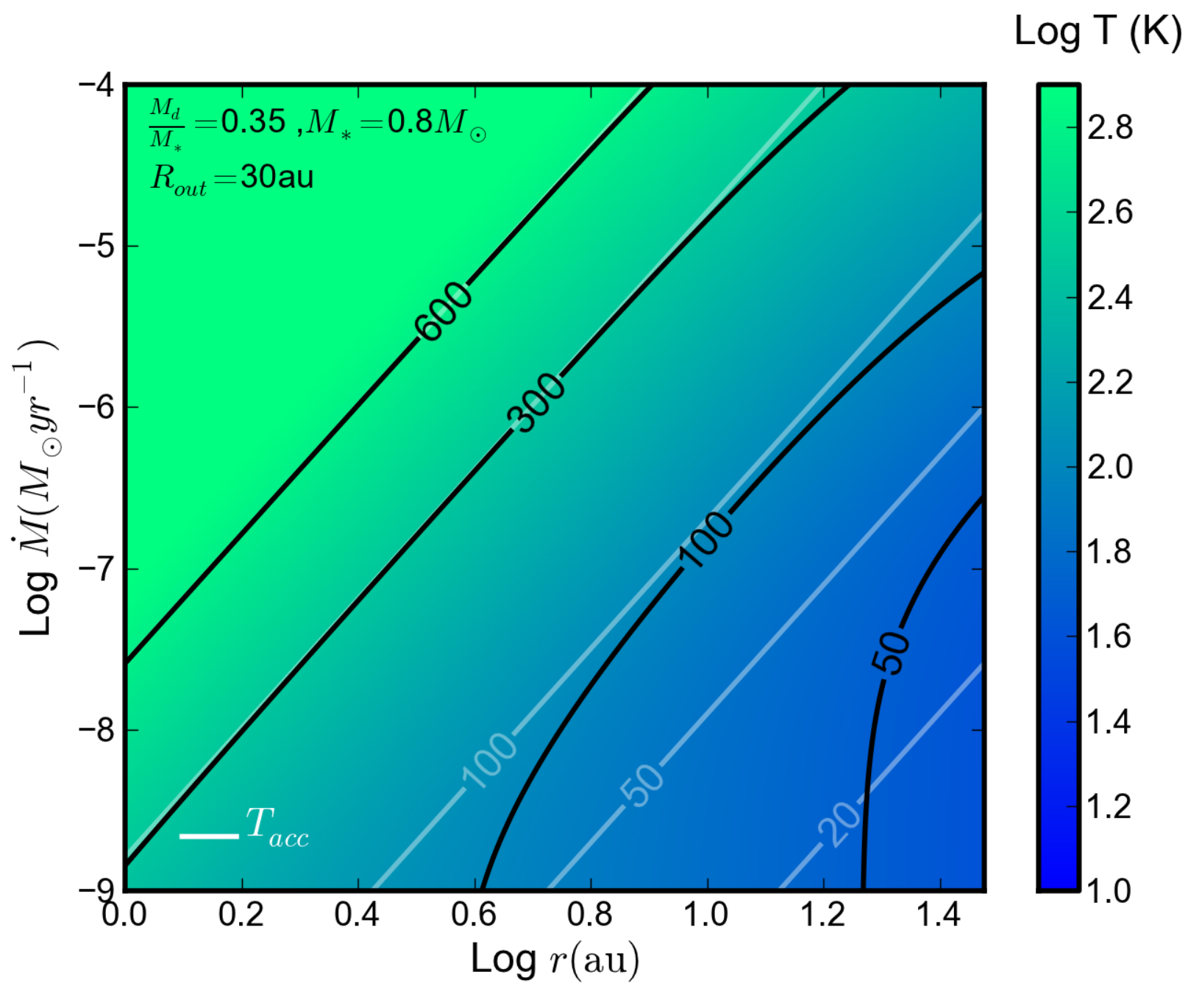}
\end{minipage}%
\begin{minipage}{.62\textwidth}
\centering
\hspace{-0.8in}
\includegraphics[width=3.3in,left]{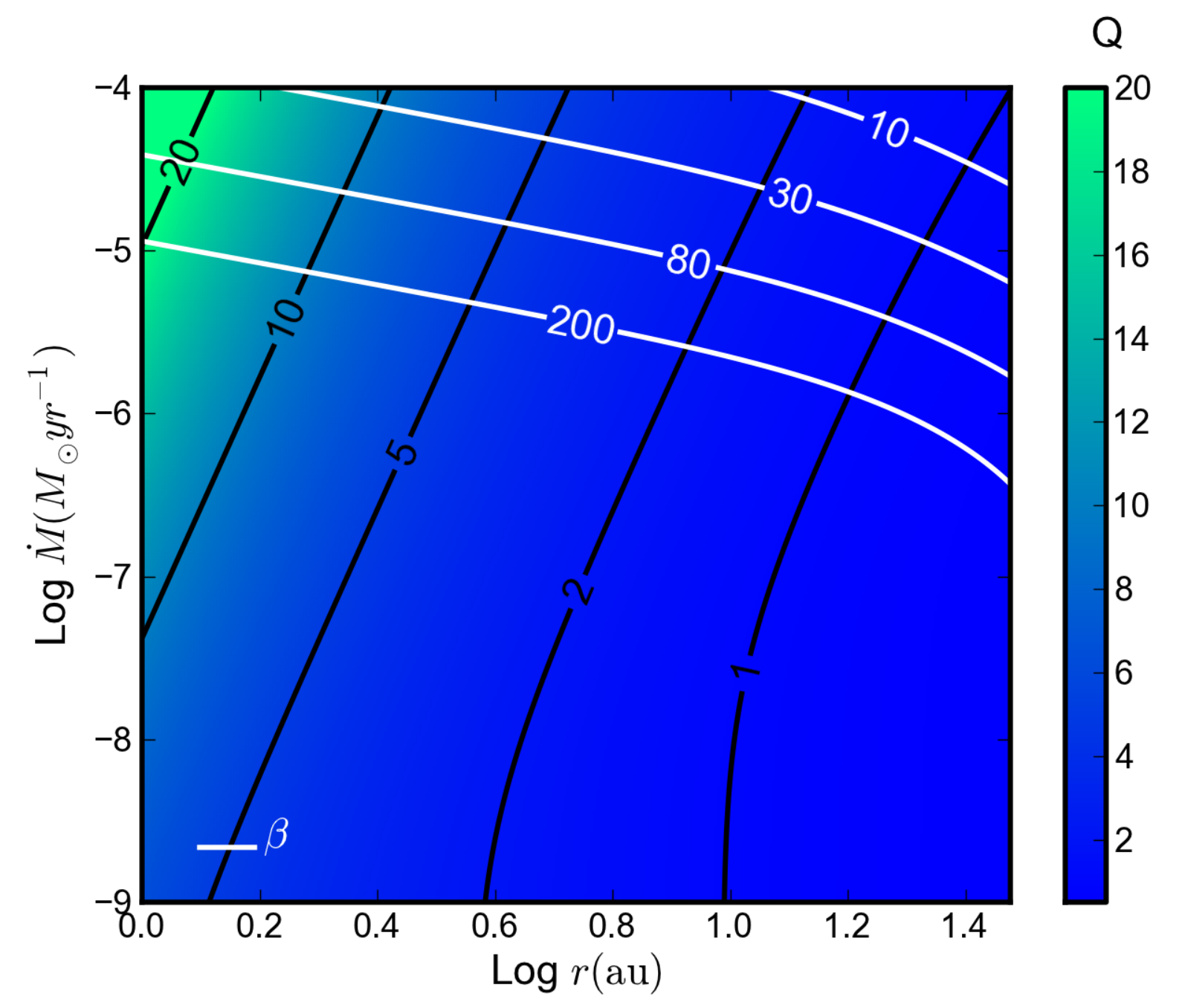}
\end{minipage}%
\end{figure}
\begin{figure}
\begin{minipage}{.62\textwidth}
\centering
\hspace{-3.1in}
\includegraphics[width=3.3in,left]{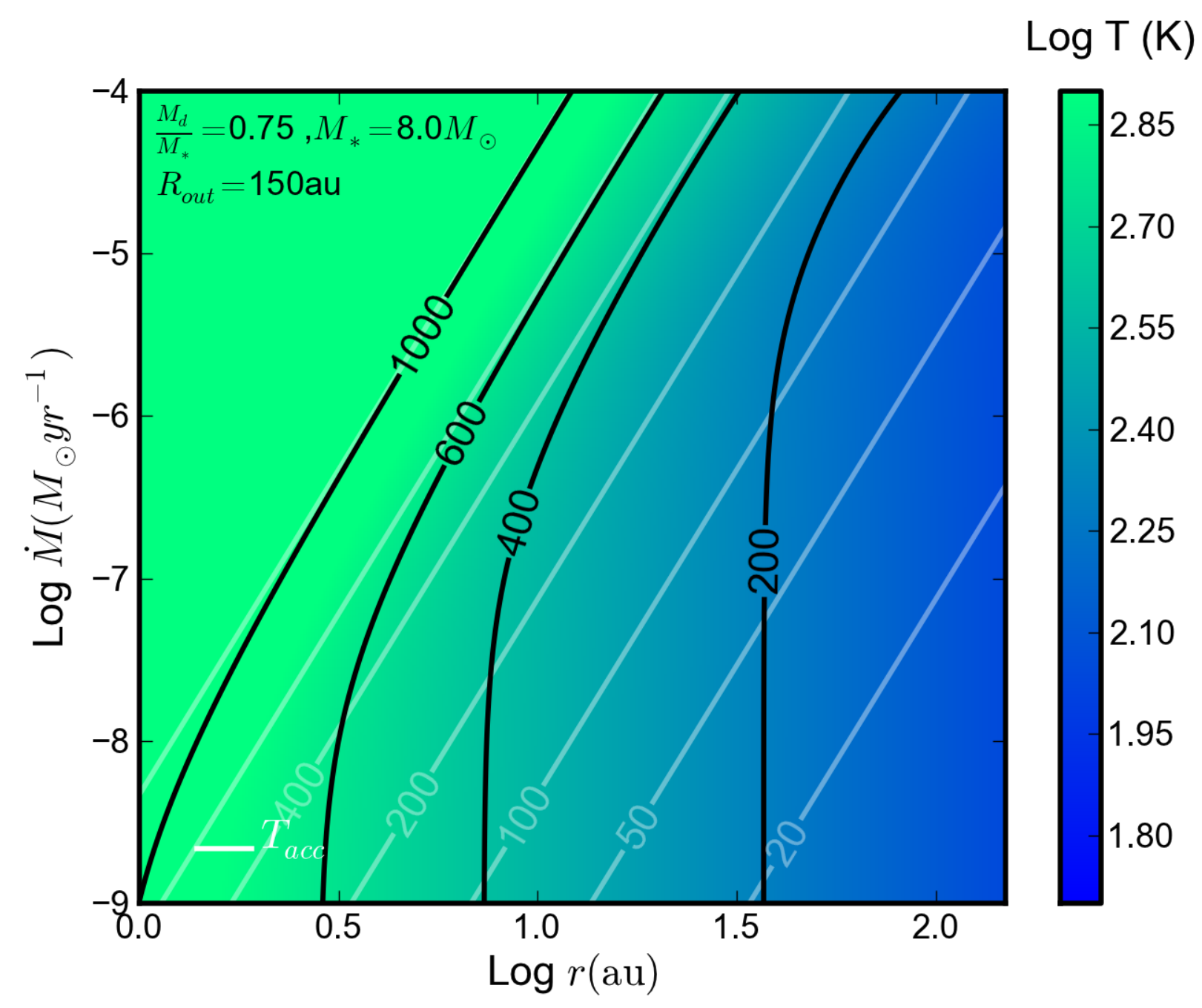}
\end{minipage}%
\begin{minipage}{.62\textwidth}
\centering
\hspace{-2.3in}
\includegraphics[width=3.3in,left]{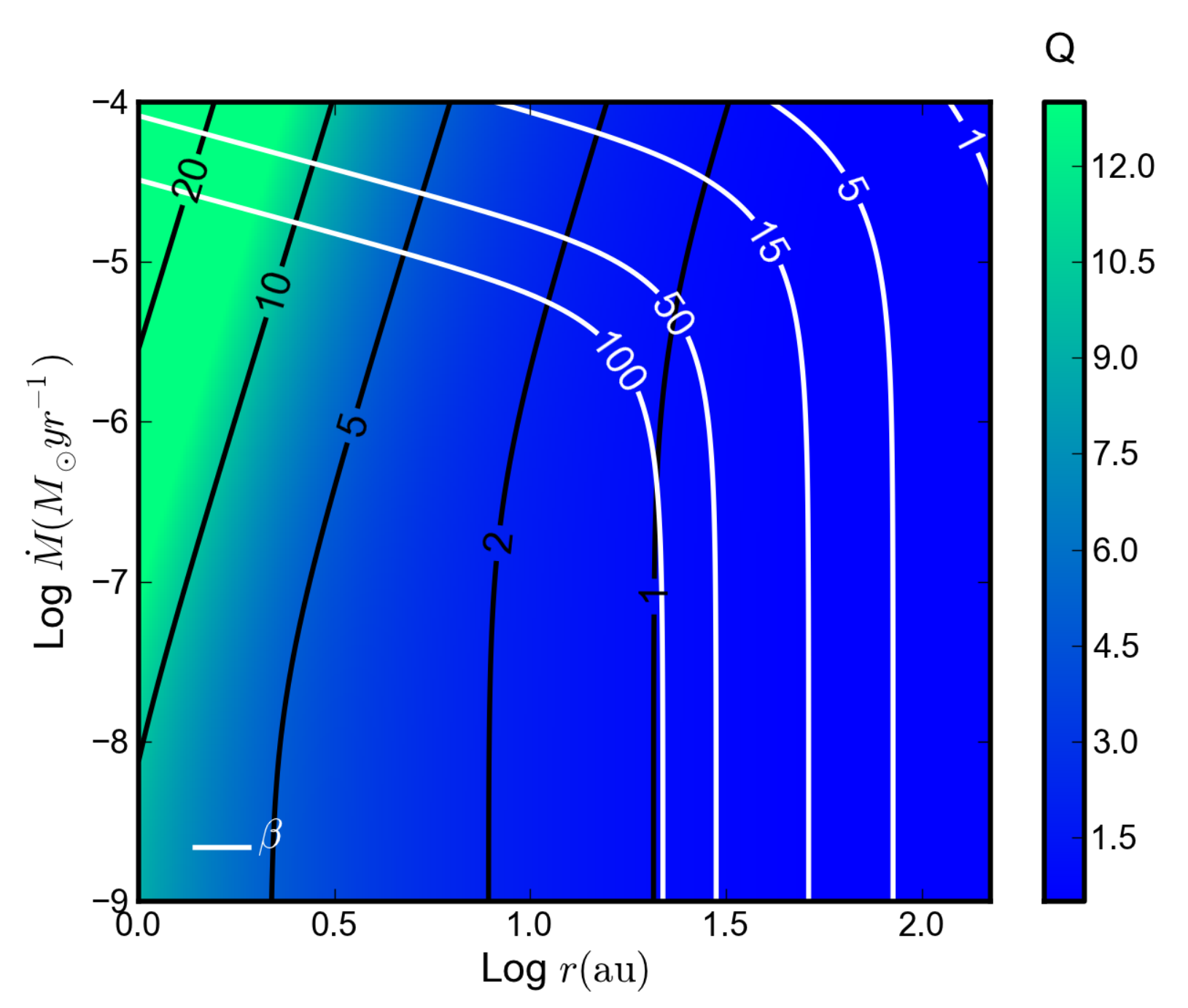}
\end{minipage}%
\caption{Three case studies for protostellar disk properties as functions of accretion rate and disk radius. The left panels show the disk temperature including (black) and excluding (white) stellar irradiaiton. The right panels show the value of $Q$ (background color and black contours) and $\beta$ (white contours). The stellar masses and disk masses used are shown in the upper left corner of the lefthand panels. For Disk 1, note that the shaded gray contours indicate solutions which require $\alpha>1$ and thus are not physical.}
\label{fig-diskstates}
\end{figure}
\subsection{The instantaneous disk state}
We demonstrate how GI manifests in protostellar and protoplanetary disks with three case studies shown in \Fig{fig-diskstates} that cover a range of parameters. Note we have not yet specified how disks might evolve into and out of these parameters, and some areas of the parameters space are unrealistic.
For each case study we provide two contour plots of fundamental disk parameters as a function of disk radius and accretion rate. We fix the disk mass, outermost radius and host star properties, and then calculate the implied properties of the disk at all locations for a range of accretion rates. 

The left hand panels show contours of the disk temperature with all three heating terms from \Eq{eq-disktemp}. For comparison we also show in white contours the temperatures that would be derived neglecting stellar irradiation, as this has been omitted in many numerical simulations. The accretion rate not only contributes to the viscous heating term, but also the stellar heating term, where for low mass stars we use the models of \cite{Baraffe:2015} to estimate the stellar radius at 1Myr.
The right hand panels show contours of $Q$ and $\beta$ (calculated from \Eq{eq:tcooldisk}). When relevant, we demarcate the boundary where the accretion rate assumed would require $\alpha>1$. This region is inconsistent with GI as a transport mechanism.\footnote{Note that the slope of the constant $\alpha$ contours, indicated by the boundary of $\alpha>1$ shaded region, are close to parallel with lines of constant $\Mdot$. This suggests that the chosen temperature and density profiles are reasonably consistent with steady-state, constant $\alpha$ disks.}  We describe each of the three cases in detail below:

\paragraph{Disk 1}:
In \Fig{fig-diskstates}a we show results for $M_d/M_* = 0.05, R_{\rm out} = 100{\rm au}$, around a solar mass star. This disk is most consistent with observations of Class II protoplanetary disks. It is immediately clear that these disks will not be subject to GI at any radius. $Q$ is lowest at the outer edge, where cooling times are also very rapid, and the temperature is controlled by stellar irradation. In order to make the outer regions of this disk unstable to GI, the disk mass must be increased by roughly a factor of 4. In this case the cooling time remains short, suggesting that such disks would fragment, but never enter the self-regulated regime, as first recognized by \cite{Rafikov:2005,ML2005}.

\paragraph{Disk 2:}
\Fig{fig-diskstates}b shows a disk modelled after observations of IRAS 16293-2422b \citep{rodriguez05}. This disk is most likely associated with an isolated Class 0 source. The disk appears to be massive and compact, consistent with a very young age, with $M_d/M_* = 0.35, R_{\rm out} = 30\rm{au}$ around a $0.8\Msun$ protostar. The \cite{rodriguez05} model suggests an accretion rate of order $10^{-6}\Msun/\rm{yr}$, but we show the full range as for the other examples. Because the disk has low $Q$ and is slowly cooling, we expect one of two outcomes depending on the rate at which mass is accreted onto the disk. Up to some critical rate, the disk can process the material via GI-induced transport and remain quasi-stable. This would require a combination of thermal regulation and mode-mode coupling. At larger infall rates, the disk will begin to grow in mass, and either revert to the first state of self-regulation, or increase in mass until it fragments.

\paragraph{Disk 3}
\Fig{fig-diskstates}c shows a disk model appropriate for a massive star, such as that identified by \cite{Shep2001}, consistent with  theoretical models of high mass star formation \citep{KM06,krumholz07}.  Here $M_{\rm d}/M_* = 0.75, R_{\rm out} = 150\rm{au}$ and $M_* = 8\Msun$. Unlike the low mass stars in the previous two examples, high mass stars are expected and observed \citep{Klaassen:2011} to have higher infall rates. Therefore the upper half of the plot is more representative of likely parameters. Here we see that the entire outer logarithmic decade in radius will have $Q\sim 1$. While at distances of the order of tens of au the cooling time is relatively long and the disk might survive in a self-regulated state, beyond $\sim 50$au the cooling time becomes short enough to induce fragmentation. The disk temperatures are higher in part because massive stars arrive on the ZAMS and start fusion while still accreting, but they still  have low $Q$ values.  The high mass ratio represented in this example may be typical because the high infall rates tend to drive up disk masses. In this case, simulations suggest that disk undergo strong $m=1$ dominated transport, and in many cases fragmentation into stellar companions \citep{ARS89,KMKK10}. The susceptibility of these disks to fragmentation is consistent with the observation that nearly all massive stars are found in binaries \citep{Sana:2010,Sana:2014}.

\subsection{Global Models for Disk Evolution}
We have laid out the instantaneous properties of protostellar and protoplanetary disk in the previous sections. However, protostellar disks do not exist in isolation, but are constantly fed material from the background core or filament. We argued in Section \ref{sec-sgphys} that infall can control the linear and non-linear development of GI. There, we simply imagined that a given disk, and considered what happened as we changed the infall rate. In order to understand what properties disks have during their evolution, star formation models are required that dictate $\Mdot_{\rm in}$ and ${\bf{\dot{J}_{\rm in}}}$. The state of the disk is controlled by competition between the infall rate and the disk-star accretion rate.

 Several authors have considered this global evolution: \cite{KMK08}, who used analytic models tied to numerical benchmarks for non-linear transport terms, \cite{VorBas07,Vorobyov:2009a}, who consider 2D models for the disk considering only transport driven by GI, \cite{Zhu:2010,Zhu:2012}, who consider long timescale semi-analytic models, and shorter timescale 2D simulations. These calculations suggest that low mass stars, as demonstrated by disk case studies (1) and (2), are susceptible to GI at early times, but stable against fragmentation. In order to push these disks into the fragmenting regime, one must invoke highly variable accretion or lower stellar illumination. Highly variable accretion can force disks like case study (2) up to the fragmentation threshold, where binary formation is the most likely outcome. Reduced irradiation is more relevant for older disks, where fragmentation into low mass objects is most likely as we discuss in the following section. These studies also suggest that burst-like behavior is probable when GI is coupled with other modes of angular momentum transport like the MRI. Finally, \cite{KMK08}, who explored the evolution of higher mass stars, found that more massive stars often evolve from the self-luminous global mode regime at early times into the irradiated, massive, and fragmenting regime at late times. The primary difference is that more massive stars are thought to experience higher mass infall rates, which more easily exceed the maximum GI accretion rate discussed in Section \ref{sec-satfail}.

To summarize, we have shown that protostellar and protoplanetary disks sample several of the regimes discussed in the previous section. Compact, Class 0 disks (see, for example \citealt{rodriguez05}), can be GI-unstable, self-luminous, and regulated via thermal saturation and mode-mode coupling.
For most other cases, a disk that reaches the threshold for gravitational instability will have a temperature controlled primarily by stellar irradiation, with short cooling times. Under these circumstances, the disk will either fragment, or the instability will saturate through mode-mode coupling, depending on whether or not the infall rate exceeds that which can be processed by strong global modes. 
 As a result, it is imperative to consider the impact of both stellar irradiation and infall in characterizing the behavior of these disks.

\section{GI and planet formation}\label{sec-sgplanet}
The non-linear phase of GI can lead to fragmentation of the disk into bound objects in orbit about the primary star. What are these objects? Beginning with \cite{ARS89}, there is a growing consensus that GI-borne objects are most likely to evolve into either brown-dwarfs or stellar binary companions around sufficiently massive stars \citep{KM06,KMCY10,Stamatellos:2009,Zhu:2012,Forgan:2013}. In contrast, \cite{Boss:1997}  argued that the direct collapse of clumps from gravitational instabilities can create massive gas giant planets (see also \citealt{Durisen:2007} and references therein). More recently, a series of papers starting with \cite{Nayakshin:2010}, have proposed that a combination of GI and tidal stripping can even produce rocky planets. Considering which disks are susceptible to GI and fragmentation reveals parameters unfavorable to the formation of planets, but favorable to the formation of more massive wide-orbit companions. In Section \ref{sec-sgprotostellar} we have seen that unstable disks are massive, which leads to large initial masses for fragments, often undergoing rapid infall, which can enhance the subsequent growth rates, and experiencing rapid angular momentum transport, which can cause rapid migration and destruction of marginally bound objects (which are the most likely to be low in mass in the first place). All of these features favor the formation of objects above the deuterium burning limit.
 
The disks best suited to forming lower mass objects are those that are unstable in the absence of continuing infall. These disks tend be less massive, and are thus colder when $Q=1$, which translates to smaller initial masses, less continued growth, and the possibility of slower migration in a more laminar disk. While these conditions are much more favorable for forming low mass objects, the formation of close-in companions $<20$au, with low mass $< 1\Mjup$, is the exception. 
 
 \subsection{Initial clump conditions}\label{sss-clumpic}
 The mass at which a GI perturbation becomes a bound fragment is the outcome of the non-linear phase of the instability, and is thus challenging to predict from simple analytic models. Although numerical simulations are now in principle well enough resolved to answer these questions, the masses evolve from this initial state quickly, and simulations become poorly resolved on a similar timescale. It is thus valuable to have analytic estimates with which to compare numerical models.

The simplest approximation is that:
\begin{equation}\label{eq-kratterm}
M_{\rm f} \approx \frac{\pi}{4} \Sigma \lambda^2 = 4\Mjup \frac{\Sigma}{\rm{50 g cm^{-2}}}\eqfrac{H/r}{0.2}^2\eqfrac{R}{\rm{50AU}}^{-2}
\end{equation}
where 
\begin{equation}
\lambda = 2 \pi H
\end{equation}
is the most unstable wavelength for the axisymmetric instability when $Q=1.$ The factor of 4 comes from assuming that only 1/2 of the unstable wavelength generates the over density. As noted by \cite{KMC11}, such a fragment would by definition be unbound at the $\Sigma$ associated with $Q=1$, because the size implied by $\lambda$ is larger than the Hill radius for the mass $M_{\rm f}$ in \Eq{eq-kratterm}. This is consistent with the instability producing non-linear perturbations.

\begin{textbox}
 \section{What is a planet?} 
 To determine whether gravitational instability makes planets, one requires a working definition of a planet. At present, there is no consensus in the community. In this chapter, we use  nuclear burning as the dividing line, although fragmentation is ignorant of this boundary. We also suggest that system architecture be considered in separating planetary systems from stellar multiples. While planets typical orbit a dominant central mass (or occassionally a tight binary), higher order stellar systems typically have hierarchical orbits.
\end{textbox}

\begin{figure}\label{fig-masscomp}
\centering
\includegraphics[width=5in]{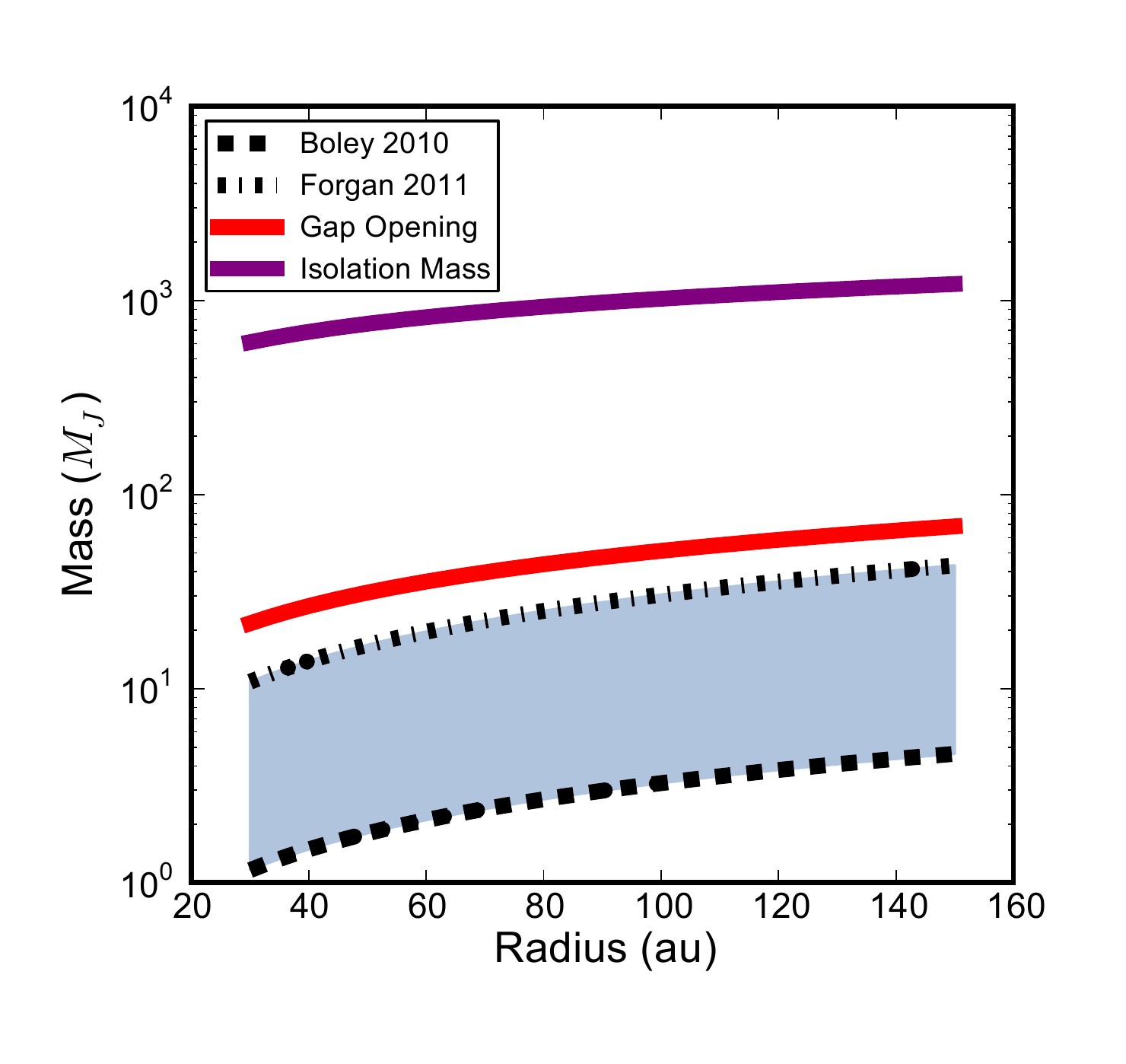}
\caption{Here we illustrate the range of initial fragment masses with lower and upper bounds set by \cite{Boley:2010} and \cite{Forgan:2011}, respectively. We also show estimates for the masses at which accretion onto the fragment from the background disk might cease: the gap opening mass and the isolation mass. See text for details about the disk model used for this calculation.}
\end{figure}

Many authors have argued for different, order unity factors in front of the dimensionally motivated $\Sigma \lambda^2$ estimate (e.g.  \citealt{Boley:2010,Forgan:2011}). In \Fig{fig-masscomp} we illustrate several different fragment mass estimates, which span an order of magnitude in mass. To compare these mass estimates we assume a temperature profile:
\begin{equation}
T = 30\rm{K}\left(\frac{r}{70\mbox{au}}\right)^{-3/7}
\end{equation}
which is (conservatively) a $\sim20\%$ reduction from the fiducial irradiation model of \citet{KMCY10}. We set the disk surface density such that $Q=1$ at every radius shown. We choose a critical value of $\beta = 13$, (relevant only for the prediction from \citealt{Forgan:2011}). In addition to the initial fragment mass estimates, we also show several limiting cases. We plot the canonical gap opening mass (using $\alpha$ set by the cooling time) and the isolation mass \citep{Lissauer:1987}. These scales are described in more detail in Section \ref{sss-growth}. Even the lowest mass estimate predicts initial fragments above $1\Mjup$ for typical disk properties. The estimates of the mass at which fragments might cease to grow are all well above the deuterium burning limit. Clearly fragments must be prevented from growing if they are to become planets.

\subsubsection{Clump Survival} \label{sss-survival}
Regardless of the mass at which clumps are born, they face a variety of threats to long term survival. These can be categorized as:
\begin{enumerate}
\item{Dispersal due to interaction with another over density (clump or spiral arm)}
\item{Tidal shredding due to slow cooling}
\item{Tidal shredding due to rapid migration}
\end{enumerate}
All three of these rely on the balance between contraction due to cooling, which allows the fragment's gravity to become ever more dominant, and disruption due to tidal forces from either the star or another overdensity in the disk. 

First consider the cooling rate of a newborn fragment. In Section \ref{sec-sgphys} we discussed the critical role of the cooling time in determining whether an unstable disk can thermally self-regulate before the instability becomes non-linear. In modelling clump survival, we must consider the cooling properties of the clump itself as it separates from the disk and increases in density. Although the numerical values of the critical cooling time are not wildly different between disk and fragment, they can in principle be different if there are sharp changes in the equation of state with density. Moreover, the appropriate radiative cooling timescale for a clump is, in fact, different from that which governs thermal self-regulation. The former is a comparison of the disk's thermal energy to the radiative cooling rate, assuming a vertically isothermal disk. The cooling timescale relevant for clump survival describes the evolution of a midplane perturbation. The appendix of \cite{KMCY10} works out this timescale as:
\begin{eqnarray}\label{eq-kmcyappendix}
t_{\rm cool} = \frac{3\gamma\Sigma c_s^2\tau}{32(\gamma-1)\sigma T^4}\times \left\{\begin{array}{cl}
\displaystyle (1-B/4)^{-1}, T >>T_0 \\
\rule{0ex}{5ex} 1, T \eqsim T_0\\
\end{array}\right.
\end{eqnarray} 
where $T_0$ is the temperature set by irradiation alone and $\kappa \propto T^B$ is the dependence of the opacity on temperature, typically $0<B<2$. Comparison with \Eq{eq:tcooldisk} shows that the perturbation cooling timescale is only different by a factor of a few at most.
The cooling rate sets the size of a fragment as a function of time. For typical fragmenting disks, which have $\beta \lesssim10$, cooling is sufficiently fast that fragments initially contract on their free fall timescale (neglecting tidal forces). 

The contracting fragment's radius must be compared with the scale on which tidal forces can shred it. Naively, one requires that:
 \begin{equation}
 r_{\rm clump} < R_{\rm H} = \eqfrac{M_{\rm clump}}{3 M_*}^{1/3}a
 \end{equation}
 where $a$ is the current semi-major axis.  Consider the origin of this limit: the Hill radius $R_{\rm H}$ determines where an object's self-gravity will dominate over tidal gravity. In the case of a collapsing fragment, pressure forces may not be negligible, weakening the inward pull of self-gravity. \cite{KMC11} find that tidal shredding can occur at radii of order $R_{\rm Hill}/3$ if pressure support is important. However, even if fragments suddenly slow their cooling, and undergo Kelvin-Helmholtz contraction rather than free fall collapse, most will contract to $<R_{\rm Hill}/3$ within a dynamical time, almost guaranteeing survival at a fixed radius.

\paragraph{Clump interaction}
Fragments may also be disrupted due to tidal interaction with other nearby fragments or spiral arms. \cite{Shlosman:1989} define a critical cooling time for clumps to collapse prior to interacting. This is relevant if the disk breaks up into multiple fragments with separations of order their own size. Since the encounter velocity of two fragments is roughly the Hill velocity $v_H = \sqrt{GM_{\rm clump}/{R_H}}$, fragments cross each others Hill radius in one dynamical time, implying that fragments must collapse to well within the Hill radius on this very rapid timescale.
This stringent requirement of collapse in one orbital period is unlikely to be necessary in fragmenting protostellar disks, which have $H/r$ sufficiently large that fragments form relatively far apart (\citealt{Boley:2010,KMKK10}, though see \citealt{Stamatellos:2009a}).  

In addition to interacting with other clumps, clumps may also interact with a passing spiral wave, since we know these should be present when fragmentation begins. \cite{Young:2016} argued that the quasi-regular nature of the spiral patterns in self-gravitating discs (as first characterized by \citealt{CLC09}) should encourage proto-fragment
disruption via collisions with spiral features over many orbital timescales. Their quantification of the intermittency of
spiral features in simulations suggested that `stochastic' fragmentation should not be possible for $\beta >$ a few $\times 10$.

\subsubsection{The physics of clump cooling}
We have outlined above some disruption mechanisms and the critical dimensionless cooling rates. But what determines the cooling rate physically? The cooling physics is nearly identical to that at the onset of star formation \citep{Larson:1969}, though self-gravitating clumps will generally be optically thick from the start \citep{Rafikov:2005}. At low temperatures, as in the background disk, cooling rates are set primarily by dust opacity. If grain growth has already commenced in such disks the opacity might be somewhat reduced compared to the ISM case. As the clump contracts under its own self-gravity, compressional heating causes the central temperatures and densities to rise. At temperatures above roughly 1200K \citep{Bell:1994, Semenov:2003}, dust sublimation removes the dominant source of opacity, accelerating cooling somewhat. Once temperatures rise all the way to 2000K, triggering molecular hydrogen dissociation, rapid collapse ensues toward planetary sizes.

 \cite{Galvagni:2012} has carried out a careful study of the cooling timescale of clumps including a complex equation of state. They extract a clump from \cite{Boley:2010} (neglecting subsequent growth, which can affect both the mass and cooling on similar timescales). They include radiative cooling due to a mixture of atomic and molecular hydrogen, with an opacity appropriate for ISM grains.
 They find that clumps that do not have their rotation artificially suppressed collapse to roughly 40\% of their initial radius within 10 dynamical times. The same clump with rotation suppressed collapses even faster to roughly 15\% of the initial radius over 10 dynamical times. This suggests that the clumps described by these models would satisfy the constraints given above, predicting survival at fixed radius.

\subsubsection{Continued Growth}\label{sss-growth}
A detailed treatment of the evolution of clumps based on their initial mass provides a useful benchmark, but omits the influence of continued accretion from the background disk.  The absolute upper limit is set by the isolation mass, which is the mass that a secondary potential embedded in the disk can attain by accreting disk material within its own Hill sphere \citep{Lissauer:1987}. The importance of continued accretion following fragment formation is likely a strong function of initial mass. If the initial mass is very large, it will clear a deep gap in the surrounding disk, possibly reducing the accretion rate. In contrast, low mass fragments, or massive fragments in rapidly accreting disks, would remain semi-embedded in the natal disk.

Gaps are opened in a disk when a relatively massive companion produces gravitational torques which repel material faster than viscosity can replenish it. Using the standard gap opening requirements of \cite{Lin:1986}  and \cite{Bryden:1999}, gaps are opened for mass ratios:
\begin{equation}\label{eq-gapcrit}
\frac{M_{\rm clump}}{M_*}> \eqfrac{H}{r}^{5/2}\sqrt\frac{3\pi\alpha}{f_g} \approx 4 \times 10^{-3} \eqfrac{\alpha}{0.1}^{1/2}\eqfrac{T}{20K}^{5/4}\eqfrac{r}{70AU}^{5/4}\eqfrac{M_*}{\Msun}^{-5/4}
\end{equation}
where $f_g=0.23$ is a geometric factor derived by \cite{Lin:1993}.
By comparing  \Eq{eq-gapcrit}, the red line, with the initial mass estimates shown in \Fig{fig-masscomp}, we see that typical fragments will be near, but below the gap opening mass, and likely able to grow at least this much. We benchmark our gap-opening estimate with a relatively large value of $\alpha$ as this is typical for strongly self-gravitating disks.  

Gap-opening, which occurs at masses well below the isolation mass, may not entirely starve growth. Numerous simulations of Type II planet migration and gap opening show that at least some material does flow across the gap, and enter into circumplanetary disks.  Indeed \cite{Duffell:2014}
find that the standard Type II migration rates that assume no mass flow are incorrect for the same reason. \cite{Lissauer:2009} found in numerical simulations that gaps roughly $5R_{\rm H}$ in width are capable of entirely shutting off accretion. Whether this limit, obtained in much lower viscosity disks, applies, is uncertain. 

\subsubsection{Migration of clumps}
Even if clumps survive at fixed radius, as discussed above, they are subject to  tidal torqueing by the parent disk. Fragments may initially be just below the gap opening mass, suggesting that they begin migrating similarly to planets undergoing type I migration. \cite{Baruteau:2011} have confirmed in 2D simulations that the migration of such objects in gravitationally unstable disks is  quite rapid, and that the gap opening criteria is accurate even in such disks. They find that despite the presence of stochastic outward forcing due to GI, the overall migration direction is inwards. Similarly, \cite{Zhu:2012} find that planets suffer divergent fates depending on whether or not they open a gap; gap-opening fragments first move in, then stall their migration, while smaller planets migrate inwards on $<10$ dynamical times. This is consistent with the work of \cite{Vorobyov:2013} in terms of migration timescale. The scenarios in which migration is most likely to stall or even reverse direction are when fragments become massive compared to the disk out of which they formed. While very interesting for understanding the distribution of brown dwarf and stellar binary companions \citep{KMKK10}, these are not relevant to the question of giant planet formation.

\subsubsection{Long timescale disruption: Tidal downsizing}
As noted at the start of this section, a second GI planet formation paradigm has arisen in the last few years wherein an initially massive fragment becomes enriched in solids and migrates inwards until its Hill Radius shrinks below the fragment radius. The envelope of the fragment is completely or partially stripped, leaving behind a low mass, rocky core with or without an atmosphere \citep{Nayakshin:2010,Boley:2011}. Successive refinements of this model include the effect of radiative feedback \citep{Nayakshin:2013}, heavy element sedimentation \citep{Nayakshin:2014}, and late stage metal enrichment \citep{Nayakshin:2015}\footnote{In \cite{Nayakshin:2015}, the neglect of heating due to the addition of solids is likely responsible for their counterintuitive finding that planetesimal accretion accelerates core collapse, which contradicts most previous work in the area \citep{Pollack:1996,Rafikov:2011}}. One significant concern with these models is the attempt to disentangle the initial disk conditions from the clump evolution. While very sensible from a modeling perspective, this can have an outsize impact on the conclusions. For example \cite{Nayakshin:2013} use artificially inserted fragments in a non-fragmenting disk as the initial conditions, which does not adequately capture the resultant disk-fragment interactions. Similarly, \cite{Nayakshin:2014} neglect accretion from the disk onto the fragment, which might fundamentally change the conclusions. These issues aside, we show in the following section that populations synthesis models of this process produce objects which are inconsistent with the rocky planet population known thus far.

\subsection{Comparison with Observed Population}
What, if any, insight can be gained from looking at the known population of planets and low mass companions? The GI model makes some predictions for the populations. The two simplest predictions come directly from the required initial conditions for clump formation: more massive stars seem more prone to developing gravitational instabilities and when clumps survive, they are prone to growing beyond the deuterium burning limit. \cite{Forgan:2013} have carried out the most extensive population synthesis calculations to date, including a  range of initial masses, accretion of solids, migration, and tidal disruption. Notably, this work excludes any gas accretion after the initial clump is formed, meaning that their final masses are a lower limit (and much of the evolution might change if mass growth occurred early). They find that roughly half of the initially formed clumps are totally destroyed, and 90\% of the remaining objects are above the Deuterium burning limit for a wide variety of initial models. A strong prediction of these models is that GI mostly makes BDs or more massive objects when it occurs, and even if tidal truncation / downsizing can in principal produce some rocky cores, it is a finely tuned part of the parameter space and unlikely to be responsible for the bulk of observed rocky planets.

\begin{figure}
 \includegraphics[width=4in]{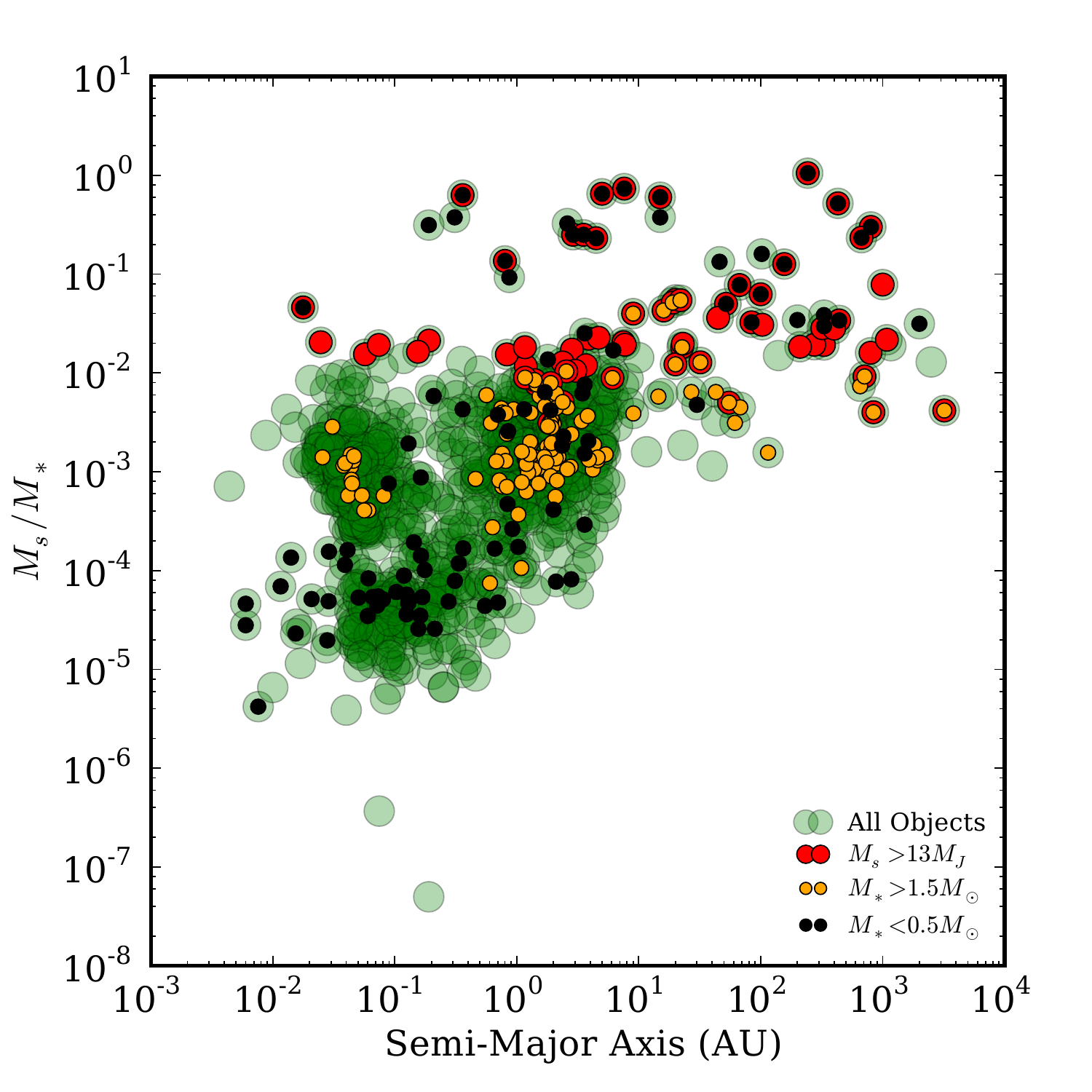}
 \caption{The distribution of planetary and other low mass companions from Exoplanets.eu as a function of mass ratio versus semi-major axis. All objects are shown in green, those above the deuterium burning limit are demarcated by red dots, those with more massive primaries are shown in yellow, while those with M-star hosts are green}
 \label{fig-exo_eu}
 \end{figure}

 
 In \Fig{fig-exo_eu}, we illustrate the population of low mass companions as compiled by the exoplanets.eu catalog. This catalog includes companions up to $30\Mjup$. We show companions as a function of mass ratio, rather than measured mass, as this has more relevance to the question of GI. Note that most of the Kepler planets are excluded from this diagram as their masses are unknown. Based on their periods and radii, they primarily occupy the region within 1AU of the host star and below $10^{-5}$ in mass ratio. GI is most likely to produce the objects at 10s of au above mass ratios of $10^{-2.5}$. In order for GI to account for more of the ``normal" planet population this outer region would likely be more populated than allowed for by current observational limits \citep{Forgan:2013}.
 
 On the opposite end of the mass spectrum, is their observational evidence for GI as a mechanism for BD and binary formation? Yes and no. On the one hand, the propensity of massive stars to binary formation via disk fragmentation is generally consistent with the abundance of binaries around more massive stars \citep{Raghavan2010,Sana:2010}. On the contrary, \cite{Kraus:2011} find that for young stars in Taurus, the lower mass objects $M_* < 0.7\Msun$ typically have  companions with smaller separations consistent with disk fragmentation ($<100AU$,) whereas stars with $0.7\Msun < M_* < 2.5\Msun$ have more companions at larger separations. Recent work on Class 0 and I binaries now finds evidence for a bimodal distribution of separations, characteristic of a disk and core mode of fragmentation \citep{Tobin:2016}. \cite{Dupuy:2011} found that the orbital distribution of BD binaries was inconsistent with ejection in pairs from fragmenting disks as predicted by \cite{Stamatellos:2009a}. As of this writing there is not a complete model for binary formation from cores and disks.  \cite{Bate:2012} provides the largest statistical sample of simulated binary formation, which matches  well with observations of the field. Unfortunately the disk fragmentation process is not sufficiently well resolved in these calculations to attribute parts of the binary distribution to one mechanism.
 
 \subsection{Role of GI in the core accretion model for planet formation}
GI may have an entirely separate means of kickstarting the planet formation process: by concentrating grains in spiral arms.  Protoplanetary disks can be well-modelled as two fluids: gas and solids. As noted in Section \ref{sec-sgphys}, while the gas feels pressure forces, which counteract gravity and slow orbital velocities, solids act as a pressureless fluid. The relative velocities between the two fluids generates drag, which is a function of particle size. This drag force results in slow radial drift of particles towards pressure maxima \citep{Whipple:1972,Adachi:1976,weiden77}. We neglect in this review the separate physics of self-gravity of solid layer in protoplanetary disks \citep{Goldreich:1973,Weidenschilling:1980}, or self-gravity of planetesimals concentrated by the streaming instability \citep{Youdin:2005}. 

 Spiral arms in a self-gravitating disk are an excellent candidate for producing pressure maxima that can enhance particle growth, much like MRI turbulence for the Streaming Instability \citep{Johansen:2009}. \citet{rice04} using 3D, global SPH simulations of gas and dust coupled via drag with a simple algorithm first demonstrated numerically that these traps  work effectively in self-gravitating disks. Later, \citet{rice06}, including the dust self-gravity,  showed that the dust sub-structure induced by the traps can become self-gravitating itself, and lead to a rapid formation of large mass rocky cores. The actual mass of these cores is related to the Jeans mass of the dust substructure, which in turn depends on the velocity dispersion induced by the gas flow in the particle distribution. Whether the velocity dispersion measured in such simulations (typically of the order of the sound speed) is physical or affected by numerics is still to be fully understood (see \citealt{booth15}). 

\citet{CL09} have raised the concern that this effect might require the cooling timescale in the disc to be relatively small based on the magnitude of the radial velocity induced by an overdensity of a given size. This would confine the effect to the outer regions of the disk. However, \cite{Gibbons:2012}, using 2D grid based simulations conclude that significant clumping occurs even for $t_{\rm cool} \approx 40\Omega^{-1}$, much above the fragmentation boundary. The particles with stopping times $\tau_s$ of order unity will typically be 1-10cm in size for the densities associated with self-gravitating disks at radii of 10-100AU. \cite{Gibbons:2014} also consider the self-gravity of the particles and their back reaction on the gas. Again, similar to the Streaming Instability \citep{Youdin:2005}, they find that particle concentrations in spiral arms can become high enough to become self-gravitating and collapse directly into large planetesimals.

A global view of particle clumping due to GI with an appropriate size distribution for the dust would be required to fully characterize the effect that GI might have in the core accretion model. This requires the use of an accurate and versatile algorithm for simulating the coupled dust and gas dynamics \citep{laibe12,laibe14,price15}.

Another issue to be fully understood is the fate of the rocky cores or planetesimals formed in a self-gravitating disk as described above. Generally, the mass of the rocky objects formed by particle concentrations within spiral arms is not expected to be high enough to directly trigger core accretion and significant further growth is required. However, such growth is significantly hampered during the self-gravitating disk phase, where the gas structure (even in the absence of drag) induces gravitational potential fluctuations that pump the eccentricity of the planetesimal swarm up to very high values, thus inhibiting significant growth \citep{BCL08}. Such an early population of planetesimals can thus produce large rocky cores only if they are retained in the disk long enough that the disk has become GI stable \citep{walmswell13}.

\section{Conclusions and Outlook}
In this review, we have provided a broad overview of the physical nature of GI, illustrating the variety of ways in which the instability manifests itself. It is first and foremost an angular momentum transport mechanism, and under some circumstances it may cause disk fragmentation, which can lead to the formation of secondary companions.  We demonstrate that protostellar disks around low mass stars may be prone to a self-regulated form of the instability at early times, if their mass ratios are high. At later times and larger radii, disk are irradiation dominated and prone to fragmentation if unstable. The resultant fragments are unlikely to be below the deuterium burning limit.  Protoplanetary disks are not expected to be unstable based on current observations. Disks around more massive stars are more likely prone to both  instability and fragmentation because rapid infall drives the disks to higher masses.

There are a handful of theoretical issues that remain unresolved. The first set of open questions relate to saturation via mode coupling.   Can mode coupling operate in tandem with thermal regulation? Under what circumstance does mode-mode coupling dominate over thermal regulation? Is there a critical disk-star mass ratio at which a transition occurs? How is the maximum mode amplitude set? Does mode coupling change under rapid infall? Since this form of saturation may dominate in protostellar disks, a detailed understanding of the mechanism is warranted. The second set of open questions relates to the critical cooling time, the nature of so-called gravito-turbulence, and stochastic fragmentation. Does GI produce a turbulent energy cascade and power spectrum akin to isotropic turbulence in the ISM? Or is gravito-turbulence observed in simulations the high resolution manifestation of small-$m$ spiral mode growth? If GI does lead to turbulence, then the stochastic fragmentation at a range of cooling times is viable. If, on the other hand, the instability does not produce rare, high amplitude density perturbations, there should be a concrete value of the cooling time at which self-regulated disks transition to fragmentation

On the observational front, much progress has been made in recent years, and ALMA is poised to make significant breakthroughs as all dishes come online. Protostellar disks are now being  imaged with high enough resolution to reveal their internal structure and morphology with unprecedented details. As mentioned above, in several cases a prominent spiral structure can be observed (for example, in the NIR scattered light images of MWC 758, \citealt{Benisty:2015}).  Several papers have anticipated the exceptional ability of ALMA to detect spiral structures induced by GI both through molecular line emission \citep{krumholz07} and through dust emission \citep{CLT10,dipierro14}. It will be possible to distinguish the large scale, open spirals expected in massive disks \citep{krumholz07} from the more tightly wound spirals expected in the low mass case \citep{dipierro14}. ALMA might even be able to detect the spatial changes in opacity due to dust trapping in spiral arms \citep{dipierro15}. 

\begin{summary}[SUMMARY POINTS]
\begin{enumerate}
\item Gravitational Instability provides efficient angular momentum transport in relatively massive protostellar disks. Evidence of this phase observationally remains sparse, but is likely to improve in the coming years with more ALMA data.
\item Based on the models presented in this review, we find that protostellar disks which achieve $Q\sim1$ often have large disk-star mass ratios and temperatures controlled by stellar irradiation. In this regime, $Q$ evolves due to changes in $\Sigma$ rather than $T$, and therefore infall is the dominant driver of instability. 
\item When fragmentation does occur, BD or stellar companion formation is more likely than planet formation. Evidence that the instability ever forms planets is currently lacking.  Uncertainties in the fragmentation boundary observed in numerical simulations have minimal impact on the likelihood of planet formation via this mechanism.
\end{enumerate}
\end{summary}

\begin{issues}[FUTURE ISSUES]
\begin{enumerate}
\item Demonstration of the relevance of the instability to realistic disks requires high resolution, high sensitivity observations with ALMA coupled with sophisticated molecular line and radiative transfer models that can more reliably ascertain disk masses.
\item In protostellar disks, perhaps the main channel by which gravitational instability saturates -- mode-mode coupling -- remains poorly understood and should be the subject of future investigation.
\item Gravito-turbulence has not convincingly been shown to behave like turbulence with a multi-scale power spectrum and energy cascade. The extent to which gravito-turbulence exhibits these characteristics is intimately tied to the issue of a critical cooling time and stochastic fragmentation. Stochastic fragmentation requires a very well sampled PDF, which may or may not occur in realistic disks.
\end{enumerate}
\end{issues}

\section*{DISCLOSURE STATEMENT}
The authors are not aware of any affiliations, memberships, funding, or financial holdings that might be perceived as 
affecting the objectivity of this review.

\section*{ACKNOWLEDGMENTS}
The authors would like to thank Phil Armitage, Cathie Clarke, and Jordan Stone for comments on earlier drafts of this article. G.L. was supported by the PRIN MIUR 2010-2011 project ``The Chemical and Dynamical Evolution of the Milky Way and Local Group Galaxies'', prot. 2010LY5N2T. K.M.K. was supported in part by the Transition Grant of the Universita' degli Studi di Milano.

\bibliography{lodato,KMK_GI}

\end{document}